\RequirePackage{fix-cm}
\documentclass[twocolumn]{svjour3}

\usepackage{graphicx}
\usepackage{amsmath}
\usepackage{amssymb}
\usepackage{makeidx}
\usepackage{enumerate}
\usepackage{textcomp}
\usepackage{verbatim}
\usepackage{stmaryrd}
\usepackage{mathabx}
\usepackage{multicol}
\usepackage{multirow}
\usepackage{color}
\usepackage{mathtools}
\usepackage{bm}
\usepackage{hhline}
\usepackage{url}
\usepackage{algorithm2e}
\usepackage{flushend}

\newcommand{\muaco}{CSP\hskip0pt+\hskip0ptMuACO\emph{sm}}
\newcommand{\condnl}{}
\newcommand{\myskip}{\noalign{\vspace{0.04cm}}}
\newcommand{\algindent}{\textcolor{white}{\_\_\_}}
\newcommand{\temp}[1]{\operatorname{\mathbf{#1}}}

\journalname{International Journal on Software Tools for Technology Transfer}

\begin{document}

\title{Exact Finite-State Machine Identification from Scenarios and Temporal Properties}

\author{Vladimir Ulyantsev$^1$ \and Igor Buzhinsky$^{1, 2}$ \and Anatoly Shalyto$^1$}
\authorrunning{V. Ulyantsev, I. Buzhinsky, A. Shalyto}

\institute{
This work was financially supported by the Government of Russian Federation, Grant 074-U01, and also partially supported by the Russian Foundation for Basic Research (RFBR), research project No.~14-07-31337 mol\_a.
We also thank Maxim Buzdalov, Daniil Chivilikhin and anonymous reviewers for useful comments.
\vspace{0.25cm}\hrule width240pt\vspace{0.2cm}
Vladimir Ulyantsev\\
ulyantsev@rain.ifmo.ru\\
~\\
Igor Buzhinsky (corresponding author)\\
igor\_buzhinsky@corp.ifmo.ru\\
~\\
Anatoly Shalyto\\
shalyto@mail.ifmo.ru\\
~\\
$^1$\;Computer Technologies Laboratory, ITMO University, St. Petersburg, Russia\\
$^2$\;Department of Electrical Engineering and Automation, Aalto University, Espoo, Finland\\
}

\date{\parbox{\linewidth}{The final publication is available at Springer via\\ \url{http://dx.doi.org/10.1007/s10009-016-0442-1}}}

\maketitle

\SetEndCharOfAlgoLine{}
\setcounter{footnote}{0}

\begin{abstract}
Finite-state models, such as finite-state machines (FSMs), aid software engineering in many ways.
They are often used in formal verification and also can serve as visual software models.
The latter application is associated with the problems of software synthesis and automatic derivation of software models from specification.
Smaller synthesized models are more general and are easier to comprehend, yet the problem of minimum FSM identification has received little attention in previous research.

This paper presents four exact methods to tackle the problem of minimum FSM identification from a set of test scenarios and a temporal specification represented in linear temporal logic.
The methods are implemented as an open-source tool.
Three of them are based on translations of the FSM identification problem to SAT or QSAT problem instances.
Accounting for temporal properties is done via counterexample prohibition.
Counterexamples are either obtained from previously identified FSMs, or based on bounded model checking.
The fourth method uses backtracking.
The proposed methods are evaluated on several case studies and on a larger number of randomly generated instances of increasing complexity.
The results show that the Iterative SAT-based method is the leader among the proposed methods.
The methods are also compared with existing inexact approaches, i.e. the ones which do not necessarily identify the minimum FSM, and these comparisons show encouraging results.
\end{abstract}

\section{Introduction}
\label{sec:introduction}
Finite-state models, such as finite-state machines, or FSMs, and deterministic finite automata (DFA), are commonly used for solving various problems arising in software engineering, such as software verification and reverse engineering.
Recently, there has been growing interest in automated FSM construction based on given specifications, which are often represented as execution traces and logs \cite{heule2013software,walkinshaw2008inferring,walkinshaw2016inferring,ohmann2014behavioral}.
Other types of data employed in model construction are temporal properties \cite{walkinshaw2008inferring,chivilikhin2014combining,ohmann2014behavioral} and invariants \cite{beschastnikh2011leveraging}.
This research direction is appealing, since inferred finite-state models can help comprehend software, reveal faults in it, facilitate model-driven development, or even serve as software.

Existing techniques, such as state merging \cite{lang1998results,walkinshaw2008inferring} and metaheuristic approaches \cite{tsarev2011finite,chivilikhin2014combining}, demonstrate acceptable performance.
However, they are almost not concerned about the size of generated models: it is not always possible to obtain the FSM with the minimum number of states, and even when it is, existing methods do not provide the proof that the found automaton is indeed the smallest possible one.
Smaller FSMs are preferred since they are easier to comprehend, to maintain, and, according to the Occam's principle, are more general, which is useful in the cases of incomplete specifications.
In particular, if FSMs are further used for test case generation \cite{chow1978testing,broy2005model}, the smaller number of states leads to more concise test suites.

In order to address this problem of constructing the smallest possible FSM, or the problem of \emph{exact FSM identification}, this paper presents four exact methods of FSM identification from test scenarios and temporal properties represented in linear temporal logic (LTL) \cite{pnueli1977temporal}.
The results of Gold \cite{gold1978complexity} and Rosner \cite{rosner1992modular} on computational complexity of other finite-state model identification problems make us believe that the considered problem is NP-hard, although no proof is provided in this paper.
The proposed methods are hence based on heuristics.
Three of them translate the problem to either the Boolean satisfiability problem (SAT) or its quantified version (QSAT).
The SAT problem has been previously used in related research: in \cite{heule2010exact} the authors learn DFA, and in \cite{ulyantsev2012extended} extended finite-state machines (EFSMs) are identified.
Conversely, the translation to QSAT has not been applied in solving such problems.
In this paper it is used for FSM construction in combination with bounded model checking \cite{biere2003bounded}, which is a form of model checking \cite{clarke1999model}, an approach in formal software verification.
The remaining method is based on backtracking.
All the methods are incorporated into an open-source tool written in Java.

Another issue, which might be important in reactive software model identification, is completeness~-- the property of having a transition for each event in each state of the identified FSM.
For example, complete FSMs are essential in sequential circuit synthesis \cite{chongstitvatana1999improving}, finite-state protocol synthesis \cite{alur2014synthesizing}, and in the IEC~61499 international industrial standard \cite{vyatkin2012iec}.
The majority of existing methods neglect this requirement, but is not the case for the proposed techniques.

The proposed methods are evaluated on case studies and randomly generated instances.
They are further compared with existing inexact approaches.
First, two of them are shown to outperform the approach from \cite{chivilikhin2014combining}.
Compared to state merging \cite{walkinshaw2008inferring}, the proposed approaches need more time, but are applicable under fewer premises (such as the absence of actions on transitions).
Finally, the comparison with symbolic bounded LTL synthesis \cite{ehlers2011unbeast,ehlers2012symbolic} suggests that the proposed methods generate notably smaller models.

The rest of the paper is organized as follows.
In Section~\ref{sec:related_work}, we examine related research.
In Section~\ref{sec:mc}, we review several key concepts from the fields of model checking and bounded model checking.
The considered problem is formally stated in Section~\ref{sec:problem_statement}.
Next, in Section~\ref{sec:methods}, we describe the contribution of the paper: the proposed FSM identification techniques.
In Section~\ref{sec:exp_eval}, we evaluate them on case study systems and random instances, and then compare them with other techniques.
Section~\ref{sec:conclusion} concludes the paper.

\section{Related work}
\label{sec:related_work}
Many previously proposed finite-state model identification methods are heuristic.
The EDSM state merging algorithm \cite{lang1998results} for constructing DFA from a number of words labeled with acceptance/rejection information was among the first ones.
State merging starts from an \emph{augmented prefix tree acceptor} (APTA), a tree-shaped automaton, and iteratively merges its states until no valid merge exists.
This algorithm serves as the basis for the method of FSM identification from execution traces and LTL safety formulae proposed in \cite{walkinshaw2008inferring}.
The authors perform a number of state merging executions (the practically efficient Blue Fringe approach is chosen) with the increasing number of negative execution traces, which are obtained as contradictions between the current FSM and LTL properties.
The validity of each merge is additionally checked against the temporal properties.
This reduces the size of the search space and thus makes the state merging procedure more efficient.

While in \cite{walkinshaw2008inferring} LTL properties are either known in advance (in the ``passive'' approach) or messaged to the FSM identification tool by its user (in the ``active'' approach), in \cite{lo2009automatic} and \cite{beschastnikh2011leveraging} they are mined from software traces or logs using predefined templates.
In \cite{lo2009automatic}, the mined temporal properties are employed to guide state merging so that they are not violated.
In \cite{beschastnikh2011leveraging}, the initially constructed compact model is iteratively refined to fulfill the temporal properties and then is additionally coerced to cancel the refinements which are redundant due to an imperfect heuristic refinement procedure.
The approach \cite{beschastnikh2011leveraging} is improved in the work \cite{ohmann2014behavioral}, which focuses on learning models whose transitions are annotated with numbers indicating resource (i.e. time or memory) consumption.
Inferring models richer than simple discrete transition systems has also been attempted in \cite{walkinshaw2016inferring}, but the idea in this work is different and does not employ temporal properties: instead, finite-state machines are enriched with numeric data classifiers learnt from traces with data values.

Another group of methods is based on metaheuristics, such as genetic algorithms \cite{mitchell1998introduction} and ant colony optimization \cite{dorigo2004ant}.
The genetic algorithm has been applied for EFSM construction in \cite{tsarev2011finite}, but the work \cite{chivilikhin2013muacosm} shows that the evolutionary algorithm based on ant colony optimization solves this problem faster.

One of the ways of finding an exact solution (i.e. the FSM with the minimum number of states conforming with the specification), apart from the naive brute-force solution enumeration, is the translation of the problem to another NP-hard problem such as SAT or the constraint satisfaction problem (CSP) and feeding the obtained set of constraints to an exact solver (a third-party tool based on heuristics).
To the best of our knowledge, all existing translation-based methods currently do not support temporal specifications.
A translation-to-SAT DFA learning method, which employs labeled examples as input data, has been proposed in \cite{heule2010exact}.
This method finds a proper coloring of the so-called \emph{consistency graph}, which determines unmergeable pairs of APTA vertices.
The paper \cite{ulyantsev2015bfs} improves this approach by adding breadth-first search (BFS) symmetry breaking predicates to narrow the search space.
Another work \cite{ulyantsev2012extended}, which is based on \cite{heule2010exact}, introduces a SAT-based method of FSM synthesis from user-prepared behavior examples, or test scenarios.
Since one of the methods proposed in our paper is based on the method from \cite{ulyantsev2012extended}, we will examine it in more detail in Section~\ref{sec:from_scenarios_only}.

Finally, the problem of identifying an FSM from both scenarios and temporal properties represented in the LTL language is solved in \cite{chivilikhin2014combining}.
The solution called \muaco~combines the use of a CSP solver and a metaheuristic search with an ant colony optimization algorithm.
More precisely, the CSP solver finds the initial solution based on scenarios only, and then it is adjusted metaheuristically to account for temporal formulae.
Thus, this approach is inexact.

There has also been a large volume of research concerning the LTL synthesis problem, wherein a reactive system compliant with given LTL properties must be derived \cite{bodik2013algorithmic}.
This problem in known to be 2EXPTIME-complete \cite{rosner1992modular} in terms of specification length.
While the majority of techniques mentioned above aim to construct a finite-state model which \emph{explains} the behavior of a software system, the LTL synthesis problem requires a software system to be \emph{constructed}.
In this case LTL properties are often easier to obtain than traces, since there is no software which can generate them.
Recent works, which attempted to solve this problem in a practically feasible way, include the approach to bound the parameters of the target system \cite{ehlers2012symbolic,finkbeiner2013bounded}, paper describing a tool which implements this idea \cite{ehlers2011unbeast}, and an approach based on incremental model refinement \cite{finkbeiner2012lazy}.

\section{Model checking and bounded model checking}
\label{sec:mc}
\begin{figure}
\centering
\includegraphics[width=3.3in]{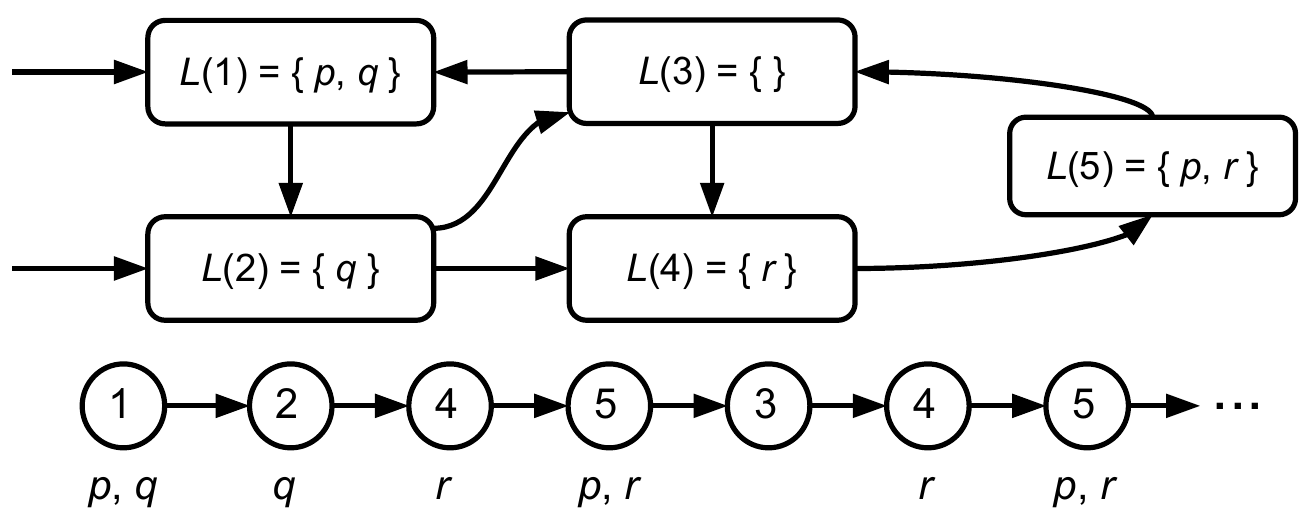}
\caption{An example of a Kripke structure (top) and an infinite path in it (bottom). The structure has 5 states, and its labeling function annotates them with three atomic propositions $p$, $q$, and $r$. Two initial states 1 and 2 are marked with incoming arrows from the left. Other arrows describe the transition relation.}
\label{fig:kripke_path}
\end{figure}

Concepts related to \emph{model checking} \cite{clarke1999model} will be extensively used in the rest of the paper.
Model checking is a formal verification technique for finite-state models which suggests describing the specification for the software in the form of \emph{temporal properties}.
One of the ways to define temporal properties is \emph{linear temporal logic} (LTL): each property to check is expressed as a formula defined over the set of infinite paths in the \emph{Kripke structure}, a special model of software execution.
A Kripke structure \cite{clarke1999model} $M$ is a quadruple $(S_K, I, T, L)$ where $S_K$ is the set of \emph{states}, $I \subset S_K$ is the set of \emph{initial states}, $T \subset S_K \times S_K$ is the \emph{transition relation}, which must be left-total (that is, from each state there is a transition to at least one state), and $L: S_K \to 2^P$ is the \emph{labeling function}, where $P$ is the set of \emph{atomic propositions}, which characterize states.
An example of a Kripke structure with an infinite path is shown in Fig.~\ref{fig:kripke_path}.

LTL formulae consist of Boolean operators ($\land$, $\lor$, $\neg$, $\rightarrow$), temporal operators and atomic propositions.
If $f$ is simply a Boolean formula, then it asserts that the first state of the path is marked with some atomic propositions and is not marked with some other ones.
If $f$ is an LTL formula, then saying that $f$ holds for a state within an infinite path means that it holds for the infinite suffix of the path starting from this state.
The following temporal operators can be used:
\begin{itemize}
\item The \emph{ne\textbf{X}t} operator: $\temp{X}f$ indicates that formula $f$ holds for the next state of the path.
\item The \emph{\textbf{G}lobal} operator: $\temp{G}f$ indicates that $f$ holds for the current state and all future states of the path.
\item The \emph{\textbf{F}uture} operator: $\temp{F}f$ indicates that there exists a future state for which $f$ holds, or it already holds for the first state of the path.
\item The \emph{\textbf{U}ntil} operator: $f\temp{U}g$ indicates that $f$ holds for a finite number of states, and then $g$ holds for the next state.
\item The \emph{\textbf{R}elease} operator: $f\temp{R}g$ indicates that either $g$ holds until both $f$ and $g$ are true in some state, or $g$ holds forever if $f$ never becomes true.
\end{itemize}

If $f$ is an LTL formula, then $M \vDash \temp{A}f$ means that $f$ is satisfied for all infinite paths in $M$ which start in $I$.
Alternatively, $M \vDash \temp{E}f$ means that there exists a path starting in $I$ for which $f$ is satisfied.

\emph{Bounded model checking} \cite{biere2003bounded}, or BMC,  is a technique to approximately verify LTL formulae by reducing the problem to a SAT instance.
The idea is to search for a counterexample for the formula being verified among finite execution paths and infinite paths with a simple structure, which enter an infinite loop at some point.
Each path, either \emph{finite} or \emph{looping}, is represented as a number of Boolean vectors $s_0, ..., s_k$, where $k$ determines the strength of the verification procedure.
This integer can be iteratively increased until a counterexample is found, or a theoretical boundary \cite{biere2003bounded} is reached which proves that BMC of the Kripke structure with the current $k$ is equivalent to its usual model checking, or the employed SAT solver fails to solve the current SAT instance.
Each $s_j \:(0 \varleq j \varleq k)$ determines the $j$-th state of the path.

\section{Problem statement}
\label{sec:problem_statement}
In this work, a \emph{finite state machine} (FSM) is a sextuple $(S, s_\mathrm{init}, E, Z, \delta, \lambda)$ where $S$ is a finite set of \emph{states}, $s_\mathrm{init} \in S$ is the \emph{initial state}, $E$ is a finite set of input \emph{events}, $Z$ is a finite set of output \emph{actions}, $\delta: S \times E \to S$ is the \emph{transition function}, and $\lambda: S \times E \to Z^*$, where $Z^*$ is the set of strings over $Z$, is the \emph{output function}.
If $\delta$ and $\lambda$ are partial functions defined over the same subset of $S \times E$, then the FSM is called \emph{incomplete}: some of its transitions are missing.
Otherwise, if $\delta$ and $\lambda$ are total functions, we call such an FSM \emph{complete}.
An FSM execution is a sequence of cycles: on each cycle the FSM receives an input event, generates an output sequence according to $\lambda$ and changes its active state according to~$\delta$.

The considered problem involves identifying an FSM with a fixed number of states $|S|$ which satisfies two types of specification: scenarios and LTL properties.
If such an FSM does not exist, this also must be eventually spotted.
Note that to find an FSM with the smallest number of states, one might try increasing $|S|$ until a solution is found.
The first type of specification is a set of \emph{test scenarios}.
A test scenario is a sequence of pairs $(e_1, A_1), ..., (e_n, A_n)$, where each $e_i \in E$ and $A_i \in Z^* \: (1 \varleq i \varleq n)$.
These pairs are called \emph{scenario elements}.

\begin{figure}
\centering
\includegraphics[width=3.1in]{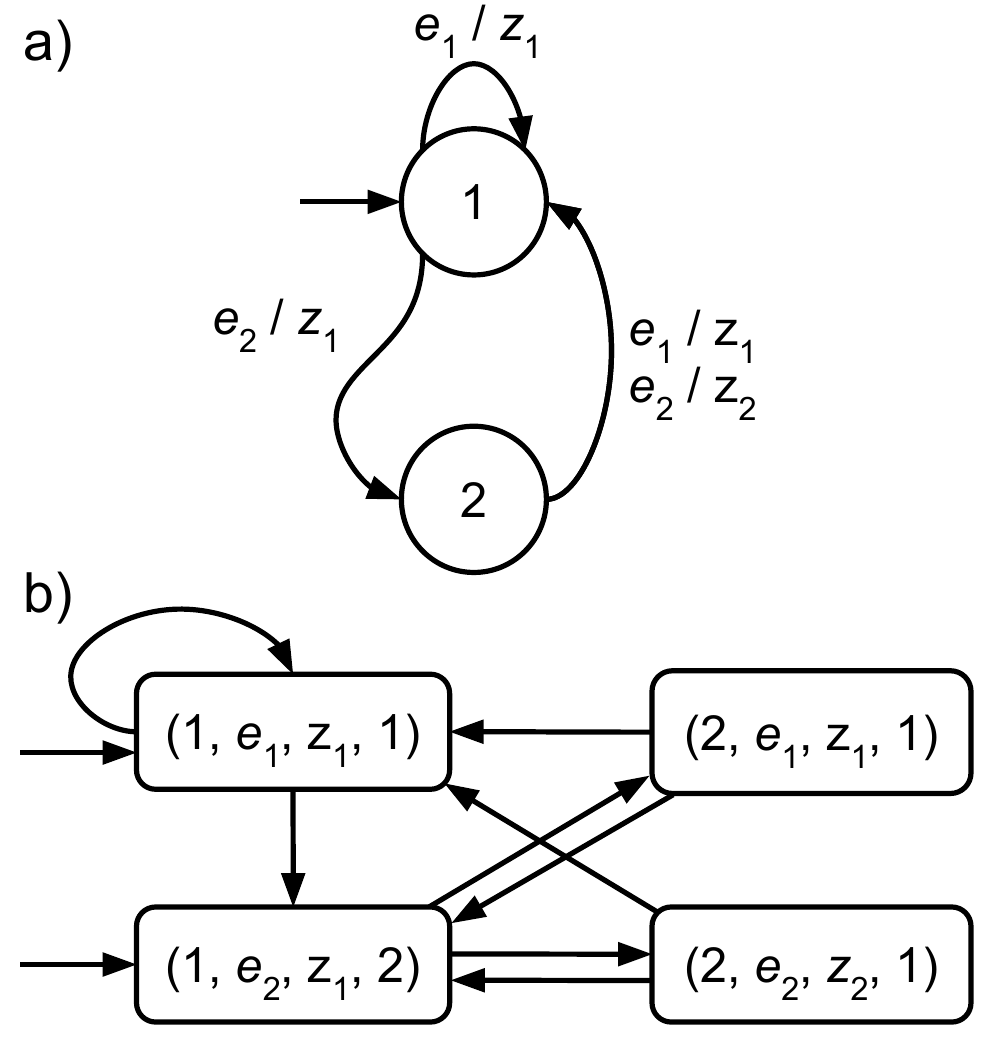}
\caption{An example of an FSM~(a) and its Kripke structure~(b).}
\label{fig:kripke}
\end{figure}

The second type of specification is a set of LTL formulae.
An FSM \emph{complies} with an LTL formula, if the formula holds for each possible execution of the FSM.
We assume the following correspondence between FSMs and their Kripke structures: the Kripke structure's states $S_K$ are FSM's transitions (thus, $S_K \subset S \times E \times Z^* \times S$), and a state of the Kripke structure is initial if and only if it corresponds to an FSM transition from its initial state $s_\mathrm{init}$:
\begin{displaymath}
I = \left\{(s_\mathrm{init}, e, \lambda(s_\mathrm{init}, e), \delta(s_\mathrm{init}, e)) \: | \: e \in E \right\}.
\end{displaymath}
Consequently, the pair composed of two states $(s_1$, $e$, $z$, $s_2)$ and $(s'_1, e', z', s'_2)$ belongs to the transition relation $T$, if and only if $s_2 = s'_1$.
An example of the described transformation is shown in Fig.~\ref{fig:kripke}.
Finally, to define the labeling function $L$, we consider the following set of atomic propositions $P$:
\begin{itemize}
\item $\mathtt{wasEvent}(e)$, $e \in E$: whether the corresponding transition of the FSM is triggered by event $e$;
\item $\mathtt{wasAction}(z)$, $z \in Z$: whether the corresponding transition of the FSM includes at least one action $z$ in its output sequence.
\end{itemize}

The second type of atomic propositions is not sufficient to express constraints which involve the positions of actions in output sequences.
Nevertheless, since these propositions were considered in one of the previous works \cite{chivilikhin2014combining} with which we compare our work, we will also use them.
Besides, generalizing $\mathtt{wasAction}$ to handle positions would not essentially influence the proposed methods.
Below are some examples of LTL formulae with the defined atomic propositions, which are satisfied for the FSM shown in Fig.~\ref{fig:kripke}:
\begin{itemize}
\item $\mathtt{wasAction}(z_1) \land \temp{X} (\mathtt{wasAction}(z_1) \lor \mathtt{wasAction}(z_2))$: the first state of the path is marked with atomic proposition $\mathtt{wasAction}(z_1)$ (and possibly with some other atomic propositions), and the second state is marked with either $\mathtt{wasAction}(z_1)$ or $\mathtt{wasAction}(z_2)$.
In terms of the corresponding FSM, this means that $z_1$ is emitted on the first cycle of FSM execution, and either $z_1$ or $z_2$ is emitted on the second cycle.
\item $\temp{G}(\mathtt{wasEvent}(e_1) \to \temp{F}(\mathtt{wasAction}(z_1)))$: each event $e_1$ received by the FSM will cause action $z_1$ in the future.
\end{itemize}

It is also possible to optionally require the identified FSM to be complete.
While describing the FSM identification techniques, we will mention the cases of both presence and absence of the completeness requirement.

The final remark in this section concerns the correspondence of our definitions of the FSM and its identification problem with the ones from previous works.
The model of EFSMs considered in \cite{ulyantsev2012extended} and \cite{chivilikhin2014combining} additionally employs guard conditions on transitions.
Such conditions depend on Boolean variables~-- an extra type of input data for an FSM.
Nevertheless, any instance of the FSM learning problem with both events and guard conditions can be transformed to an instance with events only.
Each event of the transformed instance is a pair of an event from the initial instance and a combination of variable values.
Thus, this transformation would increase the number of events in $2^{|V|}$ times, where $|V|$ is the number of variables.
For large $|V|$ such a transformation is expensive, but since in this work we deal with $|V| \varleq 2$, smarter handling of guard conditions is not considered.

Another FSM definition is the one from \cite{walkinshaw2008inferring}.
Its main difference with our one is the absence of actions, but the problem stated in \cite{walkinshaw2008inferring} assumes the optional presence of negative scenarios~-- the ones with which the identified FSM must not comply.

\section{FSM identification methods}
\label{sec:methods}
Four exact FSM identification methods are presented in this paper.
Our first method, the Iterative SAT-based approach, is largely based on a known method of identifying FSMs from test scenarios only \cite{ulyantsev2012extended} and the idea of iterative counterexample prohibition \cite{walkinshaw2008inferring}.
The second one, the QSAT-based method, uses the translation of the considered problem to QSAT and involves executing a QSAT solver.
Instead, the third approach, named the Exponential SAT-based one, executes a SAT solver on the expanded version of the quantified Boolean formula.
Eventually, the fourth and the simplest Backtracking method is based neither on SAT nor on QSAT and performs a heuristic search with backtracking.
We implemented the last method to make it the baseline in its comparison with the others.
The implementations of all the methods in Java are available online as a cross-platform software tool\footnote{\url{https://github.com/ulyantsev/EFSM-tools/}} with a command-line interface.

\subsection{Iterative SAT-based solution}
\label{sec:iterative}
The idea of the Iterative SAT-based solution is as follows.
We iteratively execute the method of identifying FSMs from test scenarios only, presented in \cite{ulyantsev2012extended}, with several adjustments.
After each iteration, we verify the obtained FSM with model checking against the LTL specification.
We employ the model checker written by the authors of \cite{tsarev2011finite} and further modified to make it output minimum counterexamples to falsified formulae.
If the FSM's Kripke structure does not comply with the specification, we prohibit the counterexamples found by the model checker using additional Boolean constraints and thus enforce the SAT solver to find a different solution after it is restarted.

An important optimization is to use the capabilities of incremental solvers \cite{een2003temporal} instead of restarts.
On each iteration, only new constraints are fed to the running instance of the solver.
This saves computation time, since the number of iterations can be large.

An approach similar to the proposed one, but based on state merging instead of the SAT problem, was introduced in \cite{walkinshaw2008inferring}.
Another work \cite{finkbeiner2012lazy} which applies related ideas is devoted to LTL synthesis.

\subsubsection{Method of identifying FSMs from scenarios only}
\label{sec:from_scenarios_only}
\begin{figure}
\centering
\includegraphics[width=3.5in]{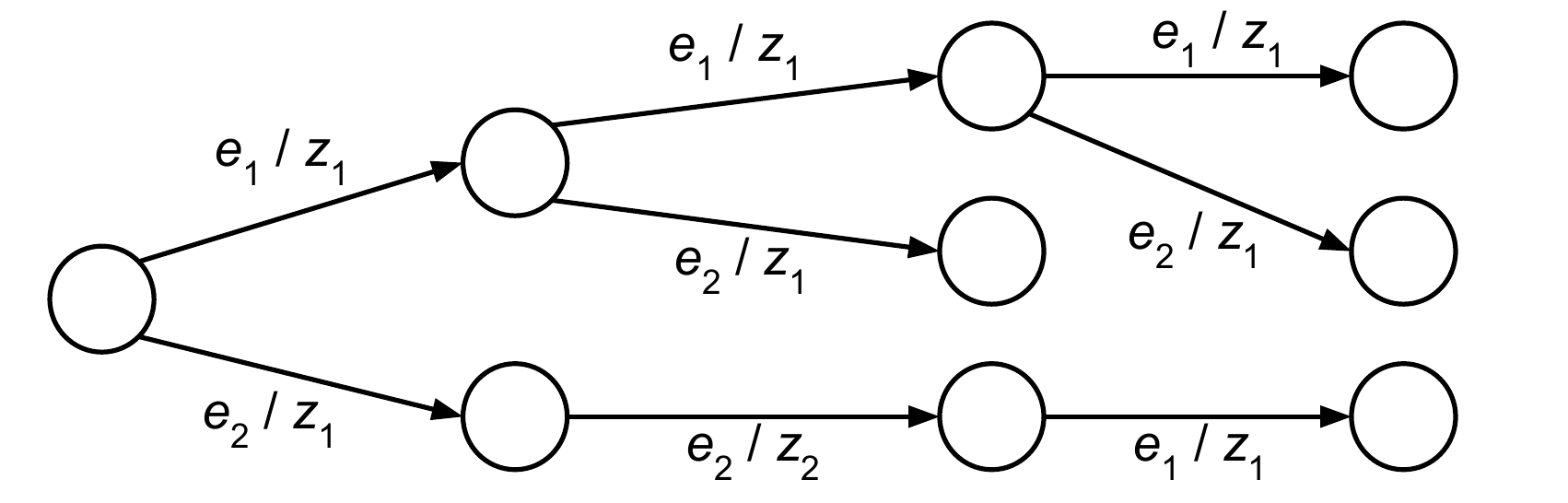}
\caption{An example of a scenario tree for four test scenarios: $(e_1, z_1), (e_1, z_1), (e_1, z_1)$; $(e_1, z_1), (e_1, z_1), (e_2, z_1)$; $(e_1, z_1), (e_2, z_1)$; and $(e_2, z_1), (e_2, z_2), (e_1, z_1)$.}
\label{fig:sctree}
\end{figure}

We now shortly describe the method from \cite{ulyantsev2012extended}.
In this method, test scenarios are merged into the \emph{scenario tree}.
An example of such tree is shown in Fig.~\ref{fig:sctree}.
Denote the set of tree nodes as $V_{\mathrm{sc}}$.
Two variable types are introduced in \cite{ulyantsev2012extended} (we slightly alter the notation from this work):
\begin{itemize}
\item $x_{v, i}$: whether node $v \in V_{\mathrm{sc}}$ of the scenario tree corresponds to state $i\:(1 \varleq i \varleq |S|)$ of the FSM ($v$ is ``colored'' into color $i$);
\item $y_{i_1, i_2, e}$: whether the transition from state $i_1\:(1 \varleq i_1 \varleq |S|)$ triggered by event $e \in E$ leads to state $i_2\:(1 \varleq i_2 \varleq |S|)$, i.e. whether $\delta(i_1, e) = i_2$.
\end{itemize}

A number of constraints enforce the proper coloring of the scenario tree and the compliance of the FSM with this tree.
Briefly, these constraints ensure that the first state of the FSM is the initial one (i.e. $x_{1, 1} = 1$), that exactly one color (FSM state) is assigned to each node of the tree, that there is no pair of inconsistent nodes \cite{heule2013software} with identical colors, that there is at most one transition $y_{i_1, i_2, e}$ in the FSM for each source state $i_1$ and event $e$, and that $y$ variables are consistent with the coloring of the tree.
Denote the logical conjunction of all these constraints as $\mathcal{S}$.

\subsubsection{Action constraints}
In \cite{ulyantsev2012extended}, output actions were not included into the SAT model, but were restored based on scenarios.
In our case, actions must be considered explicitly to facilitate counterexample prohibition (Section~\ref{sec:negsctree}).
First we need to introduce an additional variable type for output actions:
\begin{itemize}
\item $z_{i, a, e}$: whether the transition from state $i$ triggered by event $e$ produces output action $a$, i.e. whether $a \in \lambda(i, e)$.
\end{itemize}

The constraint $\mathcal{Z}$ ensures the compliance of $z$ variables with scenarios by stating that they do not contradict with each edge of the scenario tree.
Let $\mathtt{out}(v)$ be the set of outgoing edges from node $v$, then:
\begin{gather*}
\mathcal{Z} = \bigwedge\limits_{v \in V_{\mathrm{sc}}} \: \bigwedge\limits_{i = 1}^{|S|} \: \left( x_{v, i} \to \bigwedge\limits_{(e, A, v') \in \mathtt{out}(v)} \: M_{i, e, A}\right),\\
\text{where }M_{i, e, A} = \bigwedge\limits_{a \in Z} \begin{cases} z_{i, a, e}, & \text{if } a \in A\\ \neg z_{i, a, e}, & \text{if } a \notin A. \end{cases}
\end{gather*}

\subsubsection{Negative scenario tree}
\label{sec:negsctree}
We introduce the concept of the \emph{negative scenario tree}, which is used to represent counterexamples prohibited after each iteration of the method.
To do this, we need one more type of variables which will represent the colors of negative scenario tree nodes, the set of which we denote as $\overline{V}_{\mathrm{sc}}$:
\begin{itemize}
\item $\overline{x}_{v, i}$: whether node $v \in \overline{V}_{\mathrm{sc}}$ of the negative scenario tree corresponds to state $i\:\:(1 \varleq i \varleq |S|)$ of the FSM.
\end{itemize}

As in the case of the ordinary, positive scenario tree, the structure of the FSM, encoded in its Boolean model, determines the mapping between tree nodes and FSM states (this will be asserted below with Boolean constraints).
However, there are several differences between these types of trees:
\begin{itemize}
\item It is possible for negative scenario nodes to not correspond to any of the FSM states.
This is intuitive for terminal counterexample nodes, for which the opposite situation would mean that the counterexample belongs to the set of possible FSM behaviors.
Some nodes of the tree, nevertheless, correspond to FSM states: when a counterexample is added into the tree, some of its prefixes are still correct.
\item For each node of the positive scenario tree and each event, there cannot be more than one outgoing edge.
Otherwise, the tree would require the FSM to be nondeterministic.
This restriction does not apply to the negative tree: such a situation just means that more than one combination of actions is prohibited in a particular node for a particular event.
\item Generally, each counterexample to an LTL formula is an infinite path, and without loss of generality we may assume that it is composed of a finite prefix followed by a cycle \cite{clarke1999model}.
Moreover, for some formulae there are finite prefixes such that all possible infinite continuations of them are counterexamples, so we will regard such prefixes as counterexamples themselves.
A finite counterexample simply corresponds to a path from the root of the tree to the end node of the counterexample.
To represent a looping counterexample, after adding the finite prefix and a single occurrence of the cycle, a \emph{back edge} is inserted to link the end of the cycle with its beginning.
\end{itemize}

\begin{figure}
\centering
\includegraphics[width=3.35in]{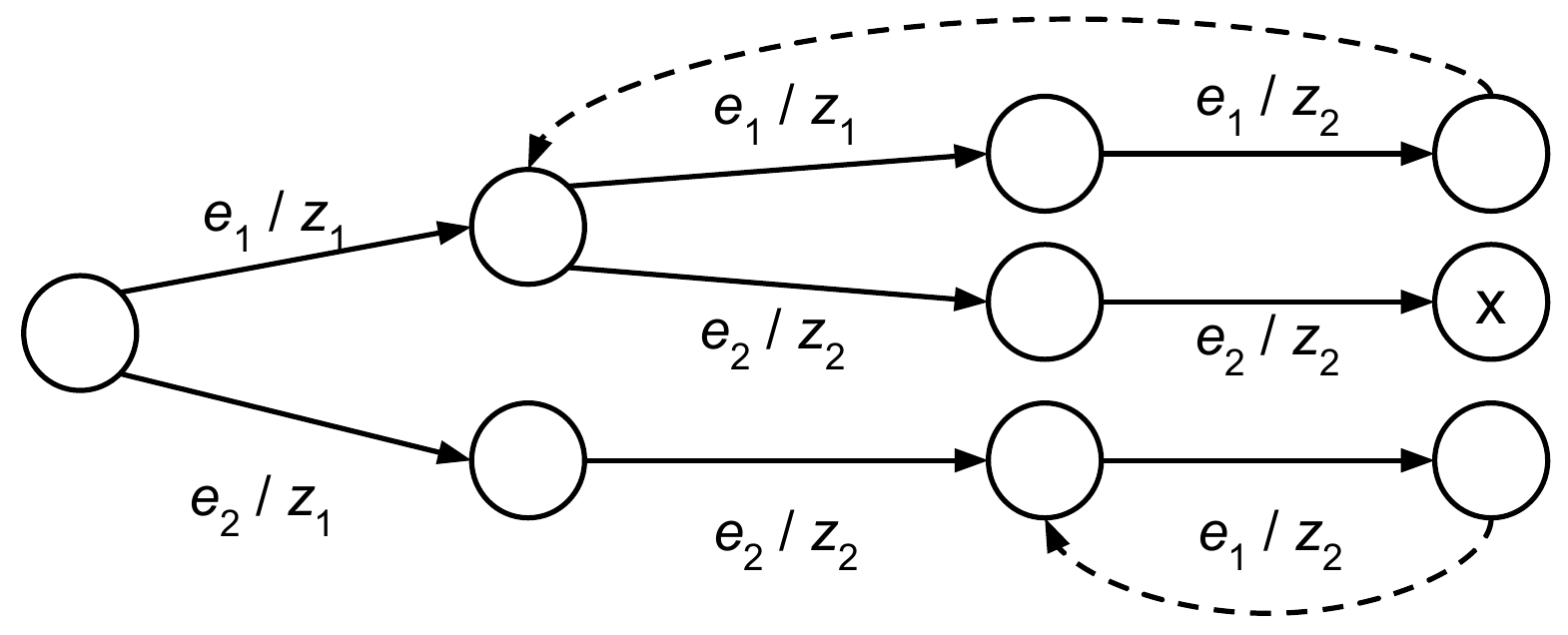}
\caption{An example of a negative scenario tree. Two looping counterexamples are $(e_1, z_1), [(e_1, z_1), (e_1, z_2)]$ and $(e_2, z_1), (e_2, z_2), [(e_1, z_2)]$ (cycles are denoted with square brackets), and the single finite counterexample is $(e_1, z_1), (e_2, z_2), (e_2, z_2)$.}
\label{fig:negsctree}
\end{figure}

An example of a negative scenario tree is shown in Fig.~\ref{fig:negsctree}.
It consists of three counterexamples: two looping ones (back edges are shown in dashed lines) and a finite one (indicated with a cross inside its end node).

Boolean constraints which specify the negative scenario tree are totally different from the ones of the positive tree.
First, proper coloring of the negative scenario tree must be ensured.
Its root (node 1) corresponds to the initial state of the FSM:
\begin{gather*}
\overline{\mathcal{S}}_1 = \overline{x}_{1, 1}.
\end{gather*}

Then, negative node colors are propagated along the edges of the tree (excluding back edges) according to the Boolean model of the FSM:
\begin{gather*}
\overline{\mathcal{S}}_2 = \bigwedge\limits_{\begin{matrix}
\scriptstyle v \in \overline{V}_{\mathrm{sc}} \\
\scriptstyle (e, A, v') \in \mathtt{out}(v) \\
\scriptstyle 1 \varleq i_1, i_2 \varleq |S|\end{matrix}} \left(\overline{x}_{v, i_1} \land y_{i_1, i_2, e} \land M_{i_1, e, A} \to \overline{x}_{v', i_2}\right).
\end{gather*}

Similarly to the positive scenario tree, it is possible to ensure that exactly one color is assigned to each negative node, but these constraints can be shown to be redundant.

Next, each added counterexample is associated with its own constraint.
The simplest case is adding finite counterexamples: their end nodes (denoted as ``terminal'' ones) are asserted to not correspond to any states:
\begin{gather*}
\overline{\mathcal{S}}_3 = \bigwedge\limits_{v \in \overline{V}_\mathrm{sc}:\; \mathtt{terminal}(v)} \: \bigwedge\limits_{i = 1}^{|S|} \neg \overline{x}_{v, i}.
\end{gather*}
Note that in general such terminal nodes may still have outgoing edges: this corresponds to situations when a shorter, more restrictive counterexample is added after a longer one.

Next, recall that a looping counterexample is added to the tree as a path consisting of the before-the-cycle prefix, a single occurrence of the cycle, and a back edge which links the end of the cycle to its beginning (see Fig.~\ref{fig:negsctree}).
In general, such an end node may still have outgoing edges due to previously added counterexamples.
The respective constraints state that cycles are forbidden: their start and end nodes (the ones linked by back edges) cannot have identical colors, i.e. they do not correspond to the same state of the FSM:
\begin{gather*}
\overline{\mathcal{S}}_4 = \bigwedge\limits_{v \in \overline{V}_\mathrm{sc}} \: \bigwedge\limits_{u:\; \mathtt{backEdge}(v, u)} \: \bigwedge\limits_{i = 1}^{|S|} \neg (\overline{x}_{v, i} \land \overline{x}_{u, i}).
\end{gather*}

Finally, the overall constraint on the negative scenario tree is denoted as $\overline{\mathcal{S}}$:
\begin{gather*}
\overline{\mathcal{S}} = \overline{\mathcal{S}}_1 \land \overline{\mathcal{S}}_2 \land \overline{\mathcal{S}}_3 \land \overline{\mathcal{S}}_4.
\end{gather*}

\subsubsection{FSM completeness}
To simply resolve the FSM completeness issue, we add the completeness constraint which ensures that for every state $i_1$ and every event $e$ there exists a transition to some state $i_2$:
\begin{displaymath}
\mathcal{C} = \mathcal{C}_\forall = \bigwedge\limits_{i_1 = 1}^{|S|} \: \bigwedge\limits_{e \in E} \: \bigvee\limits_{i_2 = 1}^{|S|} y_{i_1, i_2, e}.
\end{displaymath}
However, even when completeness is not required, we must ensure that there is at least one transition from each state of the FSM to prevent vague interpretations of LTL formulae, which are defined over infinite paths.
This can be done with a weaker constraint:
\begin{displaymath}
\mathcal{C} = \mathcal{C}_\exists= \bigwedge\limits_{i_1 = 1}^{|S|} \: \bigvee\limits_{e \in E} \: \bigvee\limits_{i_2 = 1}^{|S|} y_{i_1, i_2, e}.
\end{displaymath}

\subsubsection{Symmetry breaking}
In addition, symmetry-breaking constraints \cite{ulyantsev2015bfs} are used to speed up solver execution on unsatisfiable problem instances.
They ensure that the states of the FSM are traversed in the BFS order.
We denote them as $\mathcal{B}$.
The use of $\mathcal{B}$ requires additional variables described in \cite{ulyantsev2015bfs}, but we omit them for simplicity.

\subsubsection{Assembled formula}
Finally, we assemble all the mentioned constraints into the formula which is fed to the SAT solver:
\begin{equation}
\begin{split}
\exists \{x_{v, i},\:y_{i_1, i_2, e}, z_{i, a, e}, \overline{x}_{v, i}\}: \; \mathcal{S} \land \mathcal{Z} \land \overline{\mathcal{S}} \land \mathcal{C} \land \mathcal{B}.
\end{split}
\label{iterative_formula}
\end{equation}

The Iterative SAT-based solution is summarized in Algorithm~\ref{iter_algo}.
The function \texttt{ModelCheck} runs the model checker on the FSM and the LTL specification and returns minimum counterexamples to falsified formulae, and \texttt{SatSolve} runs a SAT solver (in the Iterative SAT-based solution, SAT solving is implemented incrementally, thus on each step only changes in the Boolean formula are fed to the solver). \texttt{SatSolve} might fail and return $null$.

\begin{algorithm}
 \SetKwData{FSM}{FSM}\SetKwData{SC}{SC}\SetKwData{LTL}{LTL}\SetKwData{f}{f}\SetKwData{counterexamples}{counterexamples}
 \SetKwFunction{ModelCheck}{ModelCheck}\SetKwFunction{SatSolve}{SatSolve}
 \KwData{set of scenarios \SC, temporal specification \LTL}
 \f $\leftarrow$ generate formula (\ref{iterative_formula})\\
 run a SAT solver in the incremental mode\\
 \While{true}{
  \FSM $\leftarrow$ \SatSolve{\f}\\
  \lIf{\FSM = null}{\KwRet{\upshape `UNSATISFIABLE'}\\}
  \counterexamples $\leftarrow$ \ModelCheck{\FSM, \LTL}\\
  \lIf{\counterexamples$\ne \varnothing$}{update $\overline{\mathcal{S}}$ within $f$\\}
  \lElse{\KwRet{\FSM}}
 }
~\\
\caption{Iterative SAT-based solution.}
\label{iter_algo}
\end{algorithm}

\subsection{QSAT-based solution}
\label{sec:qsat_based}
The QSAT-based solution employs BMC.
Assume $k$ is the BMC boundary, that is, paths with the length of $k + 1$ are checked for counterexamples.
If we find a way of identifying an FSM which satisfies scenarios and LTL properties with the boundary $k$, then we can iteratively increase $k$ until the FSM satisfies the properties in the unbounded sense (this can be checked with model checking).
Such $k$ always exists according to Theorem~1 from \cite{biere2003bounded}, and the reasons why this theorem is applicable here will become evident soon.

\subsubsection{Idea of the method}
In usual BMC, the Kripke structure to be model-checked is assumed to be known in advance.
BMC checks whether there are no paths with length bounded with $k$ in this structure for which the negation of the LTL specification hold~-- such a path would be a counterexample for the specification, which must hold for every path in the Kripke structure.
But instead of querying the SAT solver whether there exists such a path, with the help of a QSAT solver we can solve a quantified Boolean formula which states that each path in the model is not a counterexample.
Furthermore, we can now assume that the Kripke structure is not known in advance and add the existential part of the formula, which defines the structure to be identified, before the universal one, which specifies the absence of counterexamples.
The proof of the correctness of the outlined idea and its formal description are provided below.

Recall the notations $M \vDash \temp{A} f$ and $M \vDash \temp{E} f$, which state that $f$ is satisfied either for all paths or for some path in $M$ (see Section~\ref{sec:mc}).
Assume $M$ is the Kripke structure which models an FSM complying with scenarios (denote the set of such models as $M_{\mathrm{sc}}$), and for which $M \vDash \temp{A} g$, where $g$ is the LTL property we require from the FSM.
If there are several such properties, assume that $g$ is their logical conjunction.
$M \vDash \temp{A} g$ is equivalent to $\neg (M \vDash \temp{E} \neg g)$.
Next, we need to utilize two theorems from \cite{biere2003bounded}:
\begin{itemize}
\item Theorem~1. $M \vDash \temp{E} f \Leftrightarrow \exists k \vargeq 0: M \vDash_k \temp{E} f$, where the symbol ``$\vDash_k$'' denotes property satisfiability in the $k$-bounded sense.
\item Theorem~2. There is a Boolean formula $\llbracket M, f \rrbracket_k$ (defined in \cite{biere2003bounded}), which is satisfiable if and only if $M \vDash_k \temp{E} f$.
\end{itemize}

Theorem~1 implies: $M \vDash \temp{E} \neg g \: \Leftrightarrow \: \exists k = k_0 \vargeq 0: M \vDash_k \temp{E} \neg g$. Thus, if we try $k = k_0$, we need to find $M$ such that $\neg (M \vDash_k \temp{E} \neg g)$.
Then, according to Theorem~2, $M \vDash_k \temp{E} \neg g$ can be expressed as a Boolean formula $\exists p \:\: \llbracket M, f \rrbracket_k$, where $p$ is a variable assignment which determines a path in $M$ (possibly an invalid one, see clarifications below), and $f = \neg g$.
Hence, we search for $M$ such that $\exists p \:\: \llbracket M, f \rrbracket_k$ is false.
To find $M$, we start from $k = 0$ and iteratively increase it by one.
On each iteration, we solve the following quantified Boolean formula:
\begin{equation}
\label{informal_constraint}
\exists M \in M_{\mathrm{sc}} \:\: \forall p \:\: \neg \llbracket M, f \rrbracket_k.
\end{equation}

If the formula is unsatisfiable, then it is also unsatisfiable for greater $k$ (this can be inferred from the definition of $\llbracket M, f \rrbracket_k$ \cite{biere2003bounded}) and thus the desired Kripke structure $M$ does not exist.
Otherwise, we verify $M \vDash \temp{A}f$ with model checking.
If the desired Kripke structure $M$ exists, it will be found together with the corresponding FSM when $k$ reaches $k_0$.

\subsubsection{Kripke structure representation and correctness}
We have not yet discussed the way $M$ can be represented with Boolean variables and how the constraint~(\ref{informal_constraint}) can be expressed as a QSAT instance.
The translation of the stated problem to QSAT is again based on the method from \cite{ulyantsev2012extended}.
If we assume that state 1 is the initial state of the FSM, then $y$ and $z$ variables are sufficient to define the Kripke structure.
Thus, we will search both the scenario coloring determined by $x$ variables and the information sufficient to construct the Kripke structure of the FSM.
We constrain all three types of variables with $\mathcal{S}$, $\mathcal{Z}$, $\mathcal{B}$ and $\mathcal{C}$ (see Section~\ref{sec:iterative}).

\subsubsection{Path representation and correctness}
\label{sec:path_repr_corr}
We have just identified how to express $M \in M_{\mathrm{sc}}$ as a Boolean formula.
We now move to the way of defining a path in $M$ (recall that its length is $k + 1$).
We introduce the following variables for each position $j \: (0 \varleq j \varleq k)$ of the path:
\begin{itemize}
\item $\sigma_{i, j}$: the $j$-th position of the path is a transition from state $i$ of the FSM;
\item $\varepsilon_{e, j}$: the $j$-th position of the path is a transition triggered by event $e$;
\item $\zeta_{a, j}$: the $j$-th position of the path is a transition with action $a$.
\end{itemize}

\begin{figure}
\centering
\includegraphics[width=3.3in]{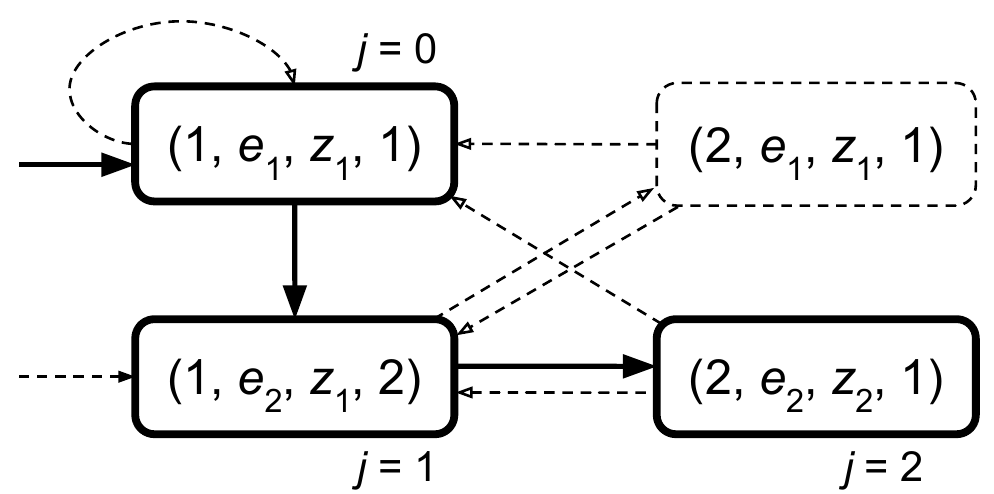}
\caption{An example of a path in a Kripke structure shown in Fig.~\ref{fig:kripke} for $k = 2$. The following path variables are true for this path: $\sigma_{1, 0}, \varepsilon_{e_1, 0}, \zeta_{z_1, 0}, \sigma_{1, 1}, \varepsilon_{e_2, 1}, \zeta_{z_1, 1}, \sigma_{2, 2}, \varepsilon_{e_2, 2}, \zeta_{z_2, 2}$, and the other ones are false.}
\label{fig:path_assignment}
\end{figure}

Thus, each Boolean vector $s_j$, introduced in the end of Section~\ref{sec:mc}, is composed of $\sigma_{i, j}\:(1 \varleq i \varleq |S|)$, $\varepsilon_{e, j}\: (e \in E)$, and $\zeta_{a, j}\:(a \in Z)$.
In fact, $\sigma$ and $\varepsilon$ variables are sufficient to determine a path in the Kripke structure, since the action sequence of the transition in the FSM can be uniquely determined from the source state of the transition and its triggering event.
Thus, $\zeta$ variables are supplemental, but later they will become helpful to express atomic propositions.
In Fig.~\ref{fig:path_assignment}, we show an example of a path in a Kripke structure with the corresponding variable assignment.

Which constraints would ensure that an assignment of the introduced variables produces a correct path?
We denote this constraint as $\llbracket M\rrbracket_k$ (from now on, some notations from \cite{biere2003bounded} are employed):
\begin{gather*}
\llbracket M\rrbracket_k = \sigma_{1, 0} \land \mathcal{P}_\sigma \land \mathcal{P}_\varepsilon \land \mathcal{P}_y \land \mathcal{P}_y^k \land \mathcal{P}_z, \text{ where }\displaybreak[0]\\\myskip\myskip
\mathcal{P}_\sigma = \bigwedge\limits_{j = 0}^k \left(\left( \bigvee\limits_{i = 1}^{|S|} \sigma_{i, j} \right) \land \bigwedge\limits_{i_1 = 1}^{|S|} \bigwedge\limits_{i_2 = i_1 + 1}^{|S|} \neg \left( \sigma_{i_1, j} \land \sigma_{i_2, j}\right) \right),\displaybreak[0]\\\myskip
\mathcal{P}_\varepsilon = \bigwedge\limits_{j = 0}^k \left(\left( \bigvee\limits_{e \in E} \varepsilon_{e, j} \right) \land \bigwedge\limits_{\{e_1 \ne e_2\}} \neg \left( \varepsilon_{e_1, j} \land \varepsilon_{e_2, j}\right) \right),\displaybreak[0]\\\myskip
\mathcal{P}_y = \bigwedge\limits_{j = 0}^{k - 1} \bigwedge\limits_{(i_1, i_2, e)} \left( \sigma_{i_1, j} \land \varepsilon_{e, j} \land \sigma_{i_2, j + 1} \to y_{i_1, i_2, e}\right),\displaybreak[0]\\
\mathcal{P}_y^k = \bigwedge\limits_{(i_1, e)} \left( \sigma_{i_1, k} \land \varepsilon_{e, k} \to \bigvee\limits_{i_2} y_{i_1, i_2, e} \right),\displaybreak[0]\\\myskip
\mathcal{P}_z =  \bigwedge\limits_{j = 0}^k \bigwedge\limits_{(i, a, e)} \left( \sigma_{i, j} \land \varepsilon_{e, j} \to \left( \zeta_{a, j} \leftrightarrow z_{i, a, e}\right)\right).
\end{gather*}

The path must start in the initial state of the FSM, therefore we need $\sigma_{1, 0}$.
The constraints $\mathcal{P}_\sigma$ and $\mathcal{P}_\varepsilon$ check that each transition in the path starts in exactly one state and is triggered by exactly one event, respectively.
Among several existing encodings of the at-most-one constraint \cite{holldobler2013sat} in $\mathcal{P}_\sigma$ and $\mathcal{P}_\varepsilon$, we have chosen the simplest binomial encoding, since these constraints do not form a significant portion of the final formula.
The constraints $\mathcal{P}_y$ and $\mathcal{P}_y^k$ (the special case of $\mathcal{P}_y$ for $j = k$) assert that the transitions in the path correspond to $y$ variables.
Note that $\mathcal{P}_y^k$ is not required if the completeness constraint $\mathcal{C}$ is included.
Finally, $\mathcal{P}_z$ defines $\zeta$ variables, enforcing correct (corresponding to $z$ variables) actions in each state of the path.

\subsubsection{Absence of a witness of the formula's negation}
\label{sec:no_witness}
By now, the only remaining thing is to express $\llbracket M, f\rrbracket_k$ as a Boolean formula.
The idea of $\llbracket M, f\rrbracket_k$ is to check whether there exists a finite or looping path in $M$ for which $f$ holds~-- its \emph{witness}.
Such path is also a counterexample for the original formula $g$.
According to \cite{biere2003bounded},
\begin{displaymath}
\llbracket M, f\rrbracket_k = \llbracket M \rrbracket_k \land \mathcal{W},
\end{displaymath}
where the path correctness condition $\llbracket M \rrbracket_k$ has been discussed previously, and $\mathcal{W}$ expresses the existence condition of a witness of $f$, the negation of $g$.
More precisely, $\mathcal{W}$ states that there exists a path in the Kripke structure on which the negation $f$ of the required LTL formula $g$ holds.
While defining $\mathcal{W}$, we will largely use the derivations from \cite{biere2003bounded}.

By $_\ell L_k \: (0 \varleq \ell \varleq k)$ we denote a Boolean formula which requires a path to be a $(k, \ell)$-loop.
In such a path, there exists a transition in the Kripke structure from the last position $k$ of the path to some position $\ell$.
$_\ell L_k$ has the following form:
\begin{displaymath}
_\ell L_k = \bigvee\limits_{(i_1, i_2, e)} \sigma_{i_1, k} \land \varepsilon_{e, k} \land \sigma_{i_2, \ell} \land y_{i_1, i_2, e}.
\end{displaymath}

Note that the looping edge is not included in the Boolean description of the path, and thus $y_{i_1, i_2, e}$ is obligatory in the definition of $_\ell L_k$.
An example of a looping path with $(2, 0)$ and $(2, 1)$-loops is the one shown in Fig.~\ref{fig:path_assignment}: its last state has transitions to the first two ones.
Next, $L_k$ will denote the existence of a $(k, \ell)$-loop for at least one $\ell$:
\begin{displaymath}
L_k = \bigvee_{\ell = 0}^k \: _\ell L_k.
\end{displaymath}

Finally, the witness condition is expressed in the following way:
\begin{displaymath}
\mathcal{W} =  \left( \neg L_k \land \llbracket f \rrbracket_k^0 \right) \lor \bigvee_{\ell = 0}^k \left( _\ell L_k \land {_\ell^{}}\llbracket f \rrbracket_k^0 \right),
\end{displaymath}
where $\llbracket f \rrbracket_k^0$ and $_\ell^{}\llbracket f \rrbracket_k^0$ are formula ``translations''~-- constraints produced from the structure of $f$.
The translations can be performed according to the rules defined in \cite{jackson2007compact}, but before this $f$ must be transformed to the negation-normal form \cite{jackson2007compact}: all negations must be propagated towards atomic propositions.
Atomic propositions encountered at position $j$ of the path are translated in the following simple way:
\begin{itemize}
\item $\mathtt{wasEvent}(e) = \varepsilon_{e, j}$;
\item $\mathtt{wasAction}(a) = \zeta_{a, j}$.
\end{itemize}

\subsubsection{Assembled formula}
We are now ready to assemble the complete quantified formula (\ref{informal_constraint}), which is further fed to a QSAT solver:
\begin{equation}
\begin{split}
\exists \{x_{v, i},\:y_{i_1, i_2, e},\:z_{i, a, e}\}: \:\: \forall \{\sigma_{i, j},\:\varepsilon_{e, j},\:\zeta_{a, j}\}: \\\myskip
\mathcal{S} \land \mathcal{Z} \land \mathcal{B} \land \mathcal{C} \land \left(\neg \llbracket M\rrbracket_k \lor \neg \mathcal{W} \right).
\label{qbf_constraint}
\end{split}
\end{equation}

Variables $x_{v, i}$, $y_{i_1, i_2, e}$, and $z_{i, a, e}$ define the FSM being identified and its Kripke structure.
Constraints $\mathcal{S}$, $\mathcal{Z}$ and $\mathcal{B}$ ensure that the assignment of these variables is valid and defines an FSM with BFS symmetry breaking predicates, and $\mathcal{C}$ guarantees the completeness of the synthesized FSM (or, if completeness is not required, it just forbids states with no outgoing transitions).
Next, each assignment of path variables either does not define a correct path ($\neg \llbracket M\rrbracket_k$) or is not a witness for $f$ ($\neg \mathcal{W}$).

\begin{figure*}
\centering
\includegraphics[height=3.2in]{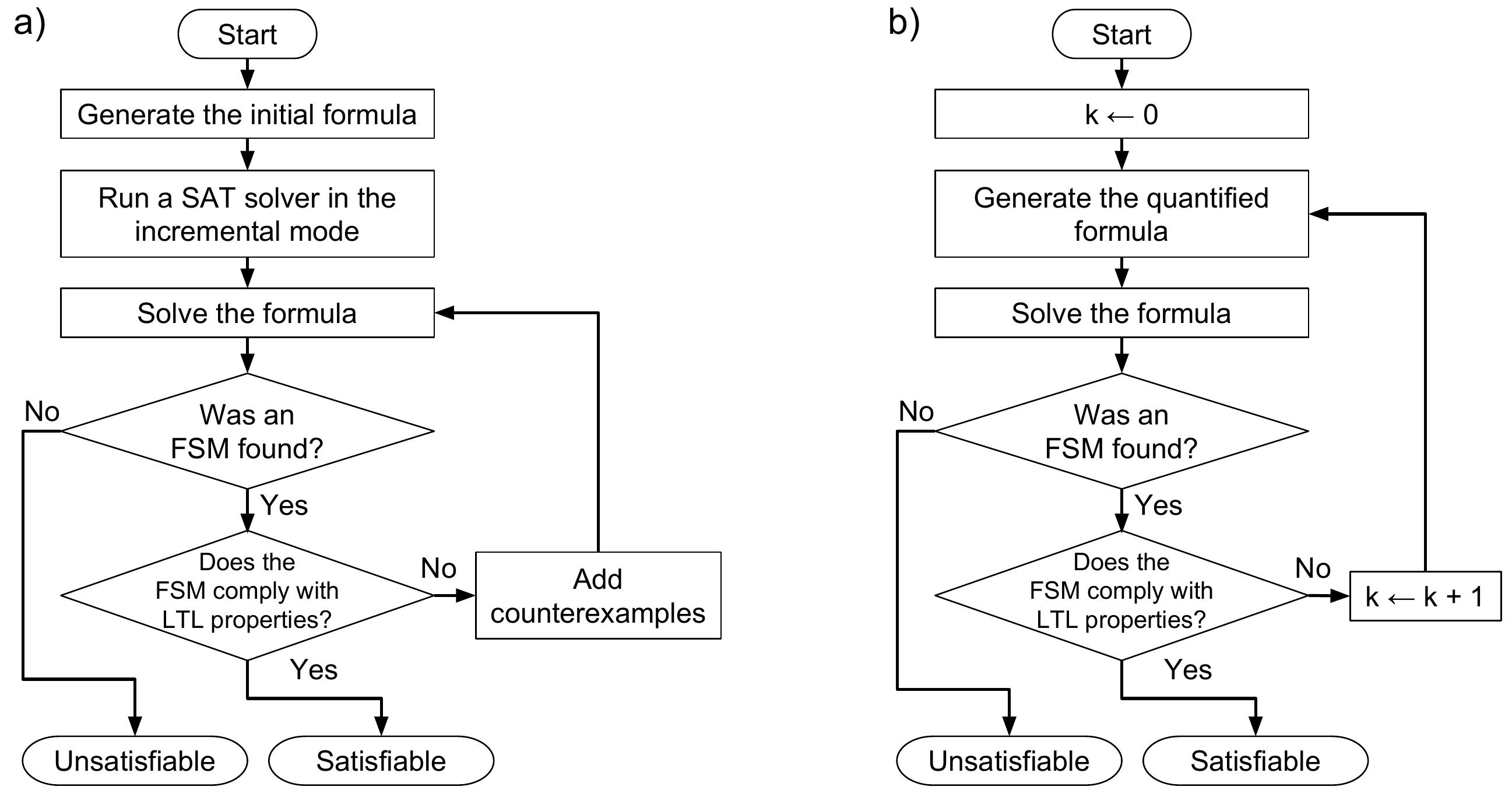}
\caption{Flowcharts of the proposed Iterative SAT-based (a) and the QSAT-based approaches (b).}
\label{fig:flowcharts}
\end{figure*}

The pseudocode of the QSAT-based solution is shown in Algorithm~\ref{qsat_algo}.
The function \texttt{QSatSolve} runs a QSAT solver to find a proper FSM, and it returns $null$ in case of the unsatisfiability of the formula.
The differences between the QSAT-based and the Iterative SAT-based solutions are also stressed in Fig.~\ref{fig:flowcharts}.
In addition, a partial example of a QSAT translation is given in Table~\ref{tab:qsat_example}.

\begin{algorithm}
 \SetKwData{FSM}{FSM}\SetKwData{SC}{SC}\SetKwData{LTL}{LTL}\SetKwData{f}{f}
 \SetKwFunction{ModelCheck}{ModelCheck}\SetKwFunction{QSatSolve}{QSatSolve}
 \KwData{set of scenarios \SC, temporal specification \LTL}
 $k \leftarrow 0$\\
 \While{true}{
  f $\leftarrow$ generate formula (\ref{qbf_constraint}), \FSM $\leftarrow$ \QSatSolve{\f}\\
  \lIf{\FSM = null}{\KwRet{\upshape `UNSATISFIABLE'}
   }\condnl
   \ElseIf{\ModelCheck{\FSM, \LTL}$= \varnothing$}{\KwRet{\FSM}}\condnl
   \lElse{$k \leftarrow k + 1$}
  }
~\\
\caption{QSAT-based solution.}
\label{qsat_algo}
\end{algorithm}

\begin{table}
\centering
\caption{Several subformulae of the QSAT translation for $|S| = |E| = |Z| = 2$, the scenario tree from Fig.~\ref{fig:sctree} and the LTL formula $\temp{G}(\mathtt{wasAction}(z_2) \to \temp{X} \mathtt{wasAction}(z_1))$. The FSM from Fig.~\ref{fig:kripke} satisfies these data. For this particular example $k = 0$ is sufficient, but $k = 1$ is used instead to make the example nontrivial.}
\begin{tabular}{cp{5.2cm}}
\hline\myskip
Name & Subformula\\\hline\myskip
Variables & $\exists x_{1..9, 1..2}, y_{1..2, 1..2, 1..2}, z_{1..2, 1..2, 1..2}$ $\forall \varepsilon_{1..2, 0..1}, \sigma_{1..2, 0..1}, \zeta_{1..2, 0..1}$\\\myskip
$\mathcal{P}_\sigma$      & $(\sigma_{1, 0} \lor \sigma_{2, 0}) \land \neg (\sigma_{1, 0} \land \sigma_{2, 0}) \: \land$\\
                          & $(\sigma_{1, 1} \lor \sigma_{2, 1}) \land \neg (\sigma_{1, 1} \land \sigma_{2, 1})$\\\myskip
$\mathcal{P}_\varepsilon$ & $(\varepsilon_{1, 0} \lor \varepsilon_{2, 0}) \land \neg (\varepsilon_{1, 0} \land \varepsilon_{2, 0}) \: \land $\\
                          & $(\varepsilon_{1, 1} \lor \varepsilon_{2, 1}) \land \neg (\varepsilon_{1, 1} \land \varepsilon_{2, 1})$\\\myskip
$\mathcal{P}_y$ & $(\sigma_{1, 0} \land \varepsilon_{1, 0} \land \sigma_{1, 1} \to y_{1, 1, 1}) \: \land $\\
                & $(\sigma_{1, 0} \land \varepsilon_{2, 0} \land \sigma_{1, 1} \to y_{1, 1, 2}) \: \land $\\
                & $(\sigma_{1, 0} \land \varepsilon_{1, 0} \land \sigma_{2, 1} \to y_{1, 2, 1}) \: \land $\\
                & $(\sigma_{1, 0} \land \varepsilon_{2, 0} \land \sigma_{2, 1} \to y_{1, 2, 2}) \: \land $\\
                & $(\sigma_{2, 0} \land \varepsilon_{1, 0} \land \sigma_{1, 1} \to y_{2, 1, 1}) \: \land $\\
                & $(\sigma_{2, 0} \land \varepsilon_{2, 0} \land \sigma_{1, 1} \to y_{2, 1, 2}) \: \land $\\
                & $(\sigma_{2, 0} \land \varepsilon_{1, 0} \land \sigma_{2, 1} \to y_{2, 2, 1}) \land ...$\\\myskip
$\mathcal{P}_z$ & $(\sigma_{1, 0} \land \varepsilon_{1, 0} \to (\zeta_{1, 0} \leftrightarrow z_{1, 1, 1})) \: \land $\\
                & $(\sigma_{1, 0} \land \varepsilon_{2, 0} \to (\zeta_{1, 0} \leftrightarrow z_{1, 1, 2})) \: \land $\\
                & $(\sigma_{1, 0} \land \varepsilon_{1, 0} \to (\zeta_{2, 0} \leftrightarrow z_{1, 2, 1})) \: \land $\\
                & $(\sigma_{1, 0} \land \varepsilon_{2, 0} \to (\zeta_{2, 0} \leftrightarrow z_{1, 2, 2})) \: \land $\\
                & $(\sigma_{2, 0} \land \varepsilon_{1, 0} \to (\zeta_{1, 0} \leftrightarrow z_{2, 1, 1})) \: \land $\\
                & $(\sigma_{2, 0} \land \varepsilon_{2, 0} \to (\zeta_{1, 0} \leftrightarrow z_{2, 1, 2})) \land ...$\\\myskip
$_0 L_1$ & $(\sigma_{1, 1} \land \varepsilon_{1, 1} \land \sigma_{1, 0} \land y_{1, 1, 1}) \: \lor $\\
         & $(\sigma_{1, 1} \land \varepsilon_{2, 1} \land \sigma_{1, 0} \land y_{1, 1, 2}) \: \lor $\\
         & $(\sigma_{1, 1} \land \varepsilon_{1, 1} \land \sigma_{2, 0} \land y_{1, 2, 1}) \: \lor $\\
         & $(\sigma_{1, 1} \land \varepsilon_{2, 1} \land \sigma_{2, 0} \land y_{1, 2, 2}) \: \lor $\\
         & $(\sigma_{2, 1} \land \varepsilon_{1, 1} \land \sigma_{1, 0} \land y_{2, 1, 1}) \: \lor $\\
         & $(\sigma_{2, 1} \land \varepsilon_{2, 1} \land \sigma_{1, 0} \land y_{2, 1, 2}) \: \lor $\\
         & $(\sigma_{2, 1} \land \varepsilon_{1, 1} \land \sigma_{2, 0} \land y_{2, 2, 1}) \: \lor $\\
         & $(\sigma_{2, 1} \land \varepsilon_{2, 1} \land \sigma_{2, 0} \land y_{2, 2, 2})$\\\myskip
$\llbracket f \rrbracket_1^0$ & $(\zeta_{2, 0} \land \neg \zeta_{1, 1}) \lor (\zeta_{2, 1} \land \text{false})$\\\myskip
$_0^{} \llbracket f \rrbracket_1^0$ & $(\zeta_{2, 0} \land \neg \zeta_{1, 1}) \lor (\zeta_{2, 1} \land \neg \zeta_{1, 0})$\\\myskip
$_1^{} \llbracket f \rrbracket_k^0$ & $(\zeta_{2, 0} \land \neg \zeta_{1, 1}) \lor (\zeta_{2, 1} \land \neg \zeta_{1, 1})$\\\myskip
\hline
\end{tabular}
\label{tab:qsat_example}
\end{table}

\subsection{Exponential SAT-based solution}
\label{sec:exp_sat_based}
Any QSAT instance can be transformed to a SAT instance by eliminating every universal quantifier: each formula $\forall x \: f$ is converted to $f\rvert_{x := 0} \land f\rvert_{x := 1}$, where the subscript expressions after vertical lines denote variable assignments.
If the formula contains $q$ universally quantified variables, then this procedure can bloat its size in up to $2^q$ times.
We take this approach in an optimized form and feed the following constraint to the SAT solver:
\begin{gather*}
\begin{split}
\exists \{x_{v, i},\:y_{i_1, i_2, e},\:z_{i, a, e}\}: \\
\mathcal{S} \land \mathcal{Z} \land \mathcal{B} \land \mathcal{C} \land \bigwedge\limits_{t \in X} \left(\neg \mathcal{P}_y \lor \neg \mathcal{P}_y^k \lor \neg \mathcal{W} \right)\Bigr\rvert_t,
\end{split}\\
\begin{matrix} \text{where } T = \left\{\{\sigma_{i, j}, \varepsilon_{e, j}, \zeta_{a, j}\} \left|\: \sigma_{1, 0} \land \mathcal{P}_\sigma \land \mathcal{P}_\varepsilon \land \mathcal{P}_z\right.\right\}\end{matrix}.
\end{gather*}

First, constraints $\mathcal{S}$, $\mathcal{Z}$, $\mathcal{B}$ and $\mathcal{C}$ do not depend on path variables and thus are included into the formula only once.
Then we iterate over the set $T$ of all valid path variable assignments~-- the ones for which $\sigma_{1, 0}$, $\mathcal{P}_\sigma$, $\mathcal{P}_\varepsilon$, and $\mathcal{P}_z$ hold.
It is important to mention that while $\sigma$ and $\varepsilon$ variables are assigned to constants (improper assignments are filtered out by $\sigma_{1, 0} \land \mathcal{P}_\sigma \land \mathcal{P}_\varepsilon$), each $\zeta_{a, j}$ is assigned to the corresponding $z_{i, a, e}$ variable, which is uniquely determined from the $\mathcal{P}_z$ constraint based on $\sigma$ and $\varepsilon$ values.
For each path variable assignment, we include the remaining part of the constraint with substituted concrete values of $\sigma_{i, j}, \varepsilon_{e, j}$, and $\zeta_{a, j}$ into the final Boolean formula.

\subsection{Backtracking solution}
\label{sec:backtracking}
\begin{figure}
\centering
\includegraphics[width=3.3in]{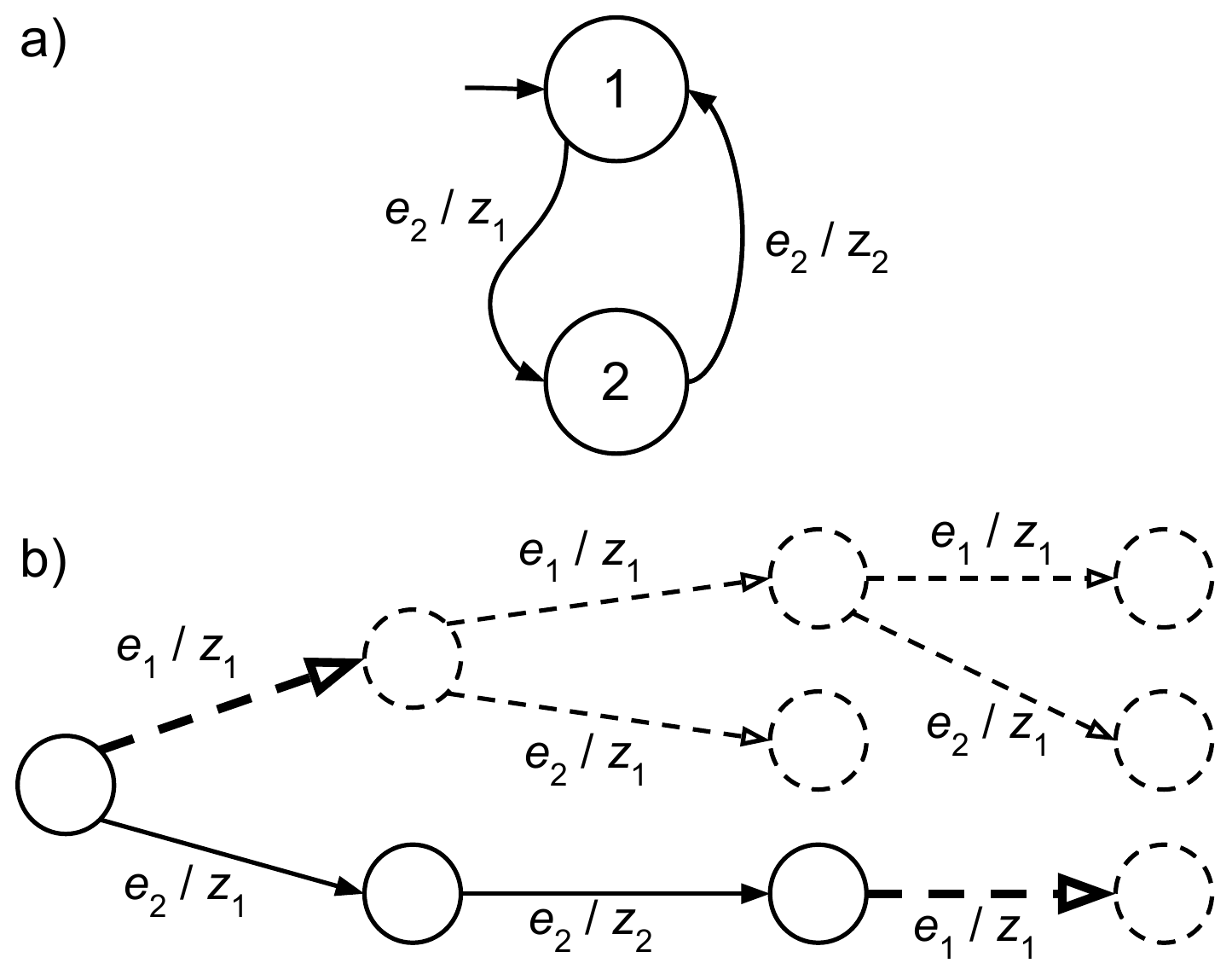}
\caption{An example of an FSM under construction (a) and a scenario tree (b) with a frontier (shown by bold dashed arrows). The frontier consists of two occurrences of the same scenario element $(e_1, z_1)$.}
\label{fig:frontier}
\end{figure}

The solution based on backtracking is the baseline one and does not involve SAT or QSAT solver execution.
A recursive procedure iterates over various (possibly incomplete) FSMs, starting from the FSM with no transitions.
It maintains the current set of edges of the scenario tree which can not yet be passed by the FSM due to the absence of transitions~-- the \emph{frontier} (see Fig.~\ref{fig:frontier} for an example).

If the frontier is not empty, then the procedure tries augmenting the FSM with one of its edges.
Each new FSM $A$ is checked for compliance with the scenario tree, and if it complies with it, then the new frontier is found.
Moreover, $A$ is verified and thus again can be rejected.
The rationale behind verifying intermediate FSMs is as follows.
If $A$ is incomplete, then the set of paths in its Kripke structure is included into the path sets of augmentations of $A$.
This allows to limit the search: if verification fails, then it will fail for all augmentations of the current FSM.
If $A$ complies with scenarios and is verified, the procedure recursively executes itself for $A$.

If the frontier is empty, we ensure FSM completeness with the same procedure as the one applied in the Iterative SAT solution, or we just return the found answer in the case of incomplete FSM identification.

Algorithm~\ref{backtracking_algo} illustrates the solution.
An additional function \texttt{FindNewFrontier} is used, which finds the frontier for the FSM augmented with a new transition, or returns $null$, if the FSM is inconsistent with scenarios.
\texttt{Backtracking} is the recursive invocation of the algorithm being defined.

\begin{algorithm}
 \SetKwData{FSM}{FSM}\SetKwData{FSMPrime}{FSM$'$}\SetKwData{SC}{SC}\SetKwData{LTL}{LTL}\SetKwData{frontier}{frontier}\SetKwData{frontierPrime}{frontier$'$}\SetKwData{trans}{edge}
 \SetKwData{source}{source}\SetKwData{dest}{destination}
 \SetKwFunction{ModelCheck}{ModelCheck}\SetKwFunction{Complete}{Complete}
 \SetKwFunction{FindNewFrontier}{FindNewFrontier}\SetKwFunction{Backtracking}{Backtracking}
 \KwData{set of scenarios \SC, temporal specification \LTL, current \FSM (initially empty), \frontier (initially contains first test elements of scenarios in \SC)}
 \trans $\leftarrow$ any edge in \frontier\\
 \For{\dest $\in 1..|S|$}{
   \lIf{\upshape $\exists$ unvisited \FSM's state $s <$~\dest\\}{\textbf{break}\\}
   \source $\leftarrow$ the state of \FSM from which \trans should \algindent be added\\
   \FSMPrime $\leftarrow$ \FSM $\cup$ transition(\source, \trans, \\ \algindent \dest)\\
   \frontierPrime $\leftarrow$ \FindNewFrontier{\SC, \FSMPrime, \frontier}\\
   \If{\upshape \frontierPrime $\ne null$ $\land$ \ModelCheck{\FSMPrime, \LTL}$= \varnothing$}{
    \uIf{\upshape \frontierPrime = $\varnothing$}{
    \uIf{\upshape completeness is required}{
     \FSMPrime $\leftarrow$ \Complete{\FSMPrime, \LTL}\\
     \lIf{\upshape \FSMPrime $\ne null$}{\KwRet{\FSMPrime}}
    }
    \lElse{\KwRet{\upshape \FSMPrime}}
    }\Else{
     \FSMPrime $\leftarrow$ \Backtracking{\SC, \LTL, \FSMPrime, \\ \algindent \frontierPrime}\\
     \If{\upshape \FSMPrime $\ne$ `UNSATISFIABLE'}{\KwRet{\upshape \FSMPrime}}
    }
   }
  }
  \KwRet{\upshape `UNSATISFIABLE'}\\
  ~\\
\caption{Backtracking solution.}
\label{backtracking_algo}
\end{algorithm}

\subsection{Preliminary comparison}
\label{sec:prel_comp}
In this subsection we present a brief preliminary qualitative comparison of the proposed methods and two other previously mentioned approaches which solve similar problems: the ones presented in \cite{walkinshaw2008inferring} and \cite{chivilikhin2014combining}.
This comparison is summarized in Table~\ref{tab:prel_comp}.

\begin{table*}
\centering
\caption{Qualitative comparison of the proposed and known FSM identification methods. Denotations: ITER~-- the Iterative SAT-based solution, QSAT~-- the QSAT-based solution, EXP~-- the Exponential SAT-based solution, BTR~-- the Backtracking solution, CMA~-- \muaco~\cite{chivilikhin2014combining}, SM~-- the passive state merging approach \cite{walkinshaw2008inferring}.}
\begin{tabular}{ccccc}
\hline\myskip
Method  & Exact & LTL class & Solver type & Formula size \\\hline\myskip
ITER    & Yes   & Arbitrary & SAT         & $O\left(l^2|S|\right)$  \\
QSAT    & Yes   & Arbitrary & QSAT        & $O\left(l^2|S| + 2^{|\mathrm{LTL}|}\right)$ \\
EXP     & Yes   & Arbitrary & SAT         & $O\left(l^2|S| + 2^{|\mathrm{LTL}| + k|S|}\right)$ \\
BTR     & Yes   & Arbitrary & --          & -- \\\hline\myskip
CMA     & No    & Arbitrary & CSP         & $O\left(l^2 + l|S|\right)$ \\
SM      & No    & Safety    & --          & -- \\
\hline
\end{tabular}
\label{tab:prel_comp}
\end{table*}

The row of the table which lists formula sizes deserves some comments.
To begin, the part of the Boolean formula expressing the compliance of the FSM with scenarios is $O(l^2|S|)$ (assuming the total length of scenarios $l \vargeq |S|$).
This scenario-related part is present in formulae generated by all three solver-based methods.
Next, LTL formula ``translations'' mentioned in Section~\ref{sec:no_witness} are in the worst case exponential of the formula size.
Overcoming this issue with the subterm extraction technique \cite{biere2003bounded} is impractical, since in our case this technique would introduce new universally quantified variables.
As we found, this slows down the QSAT solver and further increases the length of the formula produced by the Exponential SAT-based approach.
However, in some cases this estimate is polynomial: for example, for the case of a number of LTL formulae of a preassigned size, the Boolean formula length grows linearly with the number of LTL formulae.
Finally, the Exponential SAT-based solution produces the formula whose length is exponential not only of the LTL formula size, but also of $k$, since the number of considered combinations of universal variables is $O(k|S|)$.
In the estimations in this paragraph we assumed $|E|$ and $|Z|$ to be constant.

\section{Experimental evaluation}
\label{sec:exp_eval}
This section reports on experimental evaluation of the proposed approaches.
It consists of three main parts: the evaluation on case studies from the literature (Section~\ref{sec:exp_eval_case}), the evaluation on randomly generated instances (Section~\ref{sec:exp_eval_random}) and the comparison of the proposed approaches with the known ones (Section~\ref{sec:comparison}).
The experimental evaluation seeks to answer the following research questions:
\begin{itemize}
\item[]
\begin{itemize}
\item[\textbf{RQ1:}] Are the proposed methods practically applicable?
\begin{itemize}
\item[(a)] Are they applicable on FSM synthesis tasks from the literature?
\item[(b)] Is there a potential to use them in industry?
\end{itemize}
\item[\textbf{RQ2:}] To which extent are the proposed methods scalable with respect to the size of the problem?
\item[\textbf{RQ3:}] What are the benefits and shortcomings of the proposed methods in comparison with known approaches?
\end{itemize}
\end{itemize}

To answer research questions RQ1~(a), RQ2 and RQ3, three groups of experiments are conducted, which are described in Sections~\ref{sec:exp_eval_case}, \ref{sec:exp_eval_random} and \ref{sec:comparison}, respectively.
To answer RQ1~(b), all three experiments will be used.
The analysis of the results with respect to the research questions is provided in Section~\ref{sec:discussion}.

\subsection{Experimental evaluation on case studies}
\label{sec:exp_eval_case}
This subsection reports on the evaluation of the proposed FSM identification techniques on case study instances.
Its purpose is to examine whether the proposed methods are applicable in practice and how one might benefit from identifying minimum FSMs consistent with the given specification.

\subsubsection{Case study systems}
\label{sec:case_systems}
Several case study instances connected with software model inference were collected from previous works.
In the alarm clock example \cite{ulyantsev2012extended}, the FSM controlling the alarm clock must be identified.
This example is an easy one since the original set of scenarios for the clock example was very comprehensive.
Next, in the elevator example \cite{tsarev2011finite} the elevator door controlling logic is to be induced.
A more complex instance obtained from the repository\footnote{\url{https://code.google.com/p/gabp/}} of the same authors is the ATM example.
The remaining examples, adopted from \cite{walkinshaw2008inferring}, are connected not with industrial automation but with desktop software.
They include the problems of model inference for a simple text editor, the JHotDraw framework and the Jakarta Commons CVS client.
The properties of the mentioned problem instances and the number of states of reference FSMs from previous studies are summarized in Table~\ref{tab:case_examples}.

\begin{table*}
\centering
\caption{Case study systems and properties which measure their complexity.}
\begin{tabular}{cccccc}
\hline\myskip
Instance    & Events & Actions & Scenarios & LTL properties & Original $|S|$ \\\hline\myskip
Alarm clock & 16     & 7       & 38        & 11  & 3       \\
Elevator    & 5      & 3       & 9         & 13  & 5       \\
ATM         & 14     & 13      & 37        & 30  & unknown \\
Text editor & 5      & 0       & 13        & 5   & 4       \\
JHotDraw    & 6      & 0       & 27        & 10  & 7       \\
CVS client  & 16     & 0       & 12        & 29  & 18      \\\hline
\end{tabular}
\label{tab:case_examples}
\end{table*}

\subsubsection{Experiment setup}
\label{sec:case_setup}
Since the proposed methods are exact, we were interested in finding a minimum FSM for each instance.
We did it by increasing the number of states $|S|$ until the solver found a satisfying assignment.
As the starting point for $|S|$ we chose the size of the clique in the consistency graph of the scenario tree found by the greedy \emph{max-clique} algorithm \cite{heule2010exact}, which is a lower bound on $|S|_{\min}$, the optimal number of states.
Total execution times (i.e. sums of execution times for each attempted $|S|$) were recorded, and the obtained FSMs were further analyzed.
The computation was performed on the \textit{Intel Core i}7-4510{U} 2.0 GHz CPU on a single core.
Each execution series was given a time limit of 48 hours.
As for the memory limit, it was roughly equal to 8GB, the amount of memory installed on the used computer.

Boolean formulae in the context-free representation obtained as the result of translations described in Sections~\ref{sec:qsat_based} and \ref{sec:exp_sat_based} were transformed to the DIMACS format by \emph{limboole}\footnote{\url{http://fmv.jku.at/limboole/}}.
As for Boolean formulae in Section~\ref{sec:iterative}, they were sufficiently simple to generate them straightly in the DIMACS format.

To solve SAT instances, we used two solvers, both of which were highly ranked in the SAT Competition 2016\footnote{\url{http://baldur.iti.kit.edu/sat-competition-2016/}}.
In the Iterative SAT-based approach, \emph{cryptominisat}\footnote{\url{http://www.msoos.org/cryptominisat4/}}, the winner of the incremental track, was applied.
The Exponential SAT-based approach, in which incremental SAT solving was not used, employed \emph{lingeling}\footnote{\url{http://fmv.jku.at/lingeling/}}.
This solver won the third prize in the main track of the competition, but showed excellent performance on unsatisfiable instances, which is important for finding the minimum FSM.

To solve the quantified formula (\ref{qbf_constraint}) in the QSAT-based solution, we applied the \emph{DepQBF} QSAT solver \cite{lonsing2015enhancing}, which has shown the best performance among several QSAT solvers tried.
Its input format is QDIMACS, an extension of DIMACS, but \emph{limboole} can still be used to produce such inputs: it mostly remains to append quantifiers to its output.

\subsubsection{Results}
\label{sec:case_results}

The results of the evaluation are summarized in Table~\ref{tab:case_results}, which outlines whether the FSM identification methods succeeded, how much time they consumed and how large the FSMs they produced were.

\begin{table*}
\centering
\caption{Execution times (in seconds) of the proposed methods on case study systems and the state numbers of identified FSMs. ML and TL stand for a failure due to the memory limit (8GB) and the time limit (48 hours) respectively.}
\begin{tabular}{cccccc}
\hline\myskip
Instance    & ITER & QSAT    & EXP & BTR   & Minimum $|S|$\\\hline\myskip
Alarm clock & 0.7  & 77.0    & 1.6 & 0.4   & 3  \\
Elevator    & 2.2  & TL      & 5.7 & 1.3   & 5  \\
ATM         & 19.0 & TL      & ML  & 481.5 & 9  \\
Text editor & 1.3  & 14433.3 & 8.9 & 1.0   & 4  \\
JHotDraw    & 3.6  & TL      & ML  & 27.9  & 7  \\
CVS client  & TL   & TL      & ML  & TL    & 18 \\\hline
\end{tabular}
\label{tab:case_results}
\end{table*}


Unfortunately, the QSAT-based method, which has more complex theory behind it compared to other methods, failed to solve almost all instances, being successful only on the alarm clock and the text editor examples.
Its substitute, the Exponential SAT-based approach, performed better, but still violated memory limits on three instances.
The Backtracking and the Iterative SAT-based methods were much more successful, with the performance of the former being biased towards smaller instances.
Clearly, the leader in this comparison is the Iterative SAT-based method, but even it was unable to solve the CVS client instance.
More precisely, it was able to find the solution fast (in only 35 seconds) for $|S| = |S|_{\min} = 18$, but the proofs of unsatisfiability of instances with smaller $|S|$ did not terminate within the time limit.

In general, the major fraction of time is spent by the Iterative SAT-based and the Backtracking methods to prove that solutions with $|S| < |S|_{\min{}}$ (in particular, with $|S| = |S|_{\min} - 1$) do not exist.
For the Iterative SAT-based method, long delays were caused exclusively by the last iteration (after adding all counterexamples), when the SAT solver was faced with unsatisfiable problem instances.
This is the cost of solving the problem precisely.
If the proper $|S|$ is known in advance, their run times will be much smaller.

\begin{figure*}
\centering
\includegraphics[width=5.5in]{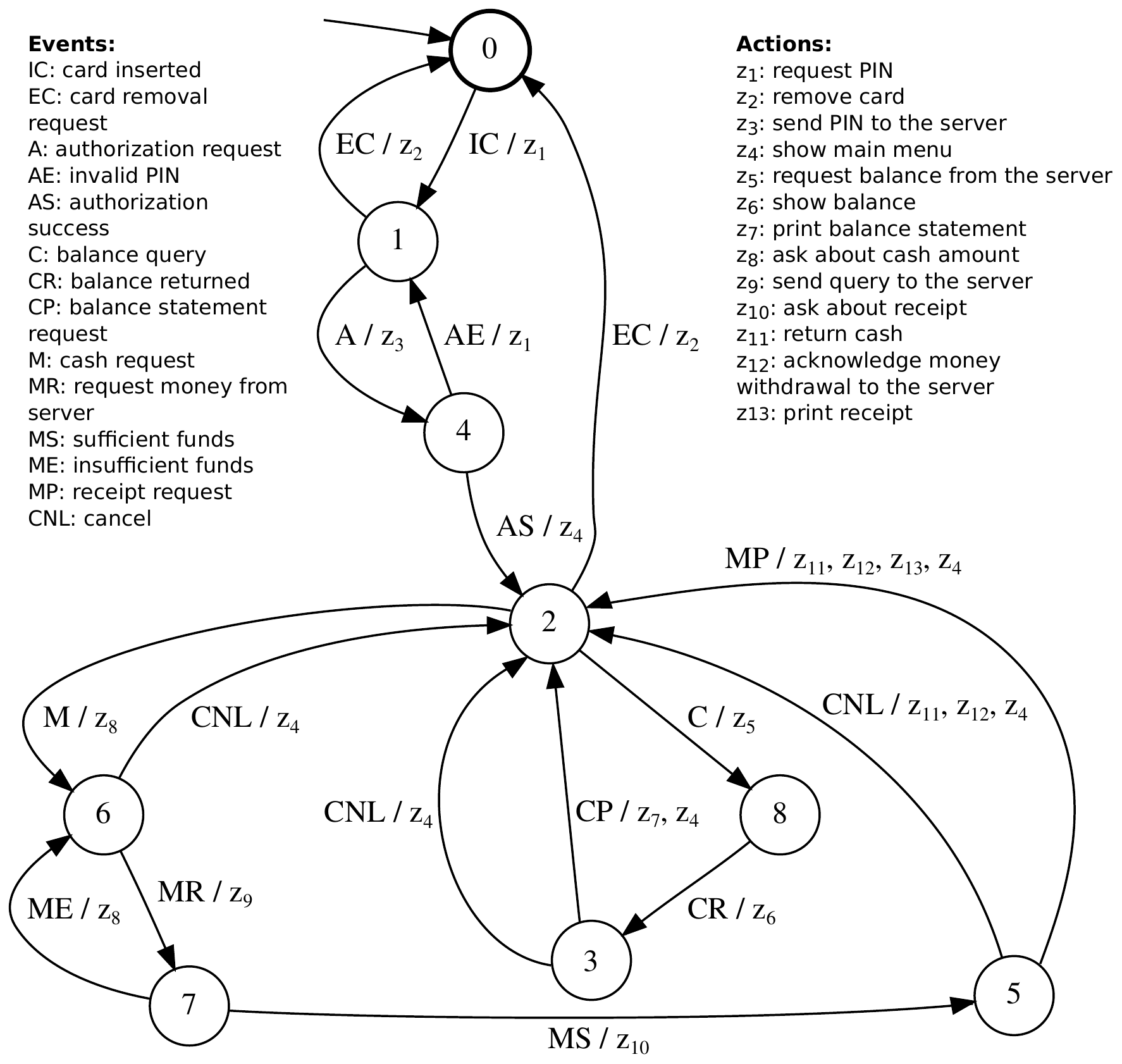}
\caption{Generated FSM for the ATM problem.}
\label{fig:app_cash}
\end{figure*}

For the presented case study instances, the target FSMs were known in advance (except the ATM example) and were proved by our methods to be minimum (except the CVS client example) by showing that there is no solution with fewer states.
The minimality of a solution implies that all its states are obligatory and meaningful, which can aid the comprehension of the system subject to model inference.
We provide the previously unknown minimum FSM for the ATM instance in Fig.~\ref{fig:app_cash}.

\subsection{Experimental evaluation on random instances}
\label{sec:exp_eval_random}
This subsection reports on the evaluation of the proposed FSM construction techniques on a larger number of random problem instances of increasing complexity.
While in the previous subsection almost all instances (except the alarm clock) specified incomplete FSM identification, in this part of the evaluation both types of instances are equally present.

\subsubsection{Instance preparation}
\label{sec:experiment_setup}
To evaluate the proposed FSM identification methods on a larger sample and on a more diverse set of LTL properties, we prepared problem instances based on random FSMs.
For each $3 \varleq |S| \varleq 12$ and both subtypes of the problem (either complete or incomplete FSM identification), 50 FSMs were randomly generated.
Each generated FSM had four events and four actions, and the lengths of output sequences on each transition were sampled uniformly at random from the set $\{0, ..., 4\}$.
Incomplete FSMs possessed 50\% of possible transitions.
Following that, four random LTL formulae were generated for each FSM with the \emph{randltl} tool \cite{duret2013manipulating}.
To ensure that the formulae were sufficiently difficult, we discarded the ones satisfied for more than 5 of 10 other random FSMs generated with the same parameters.

Then we randomly generated test scenarios by executing the generated FSMs choosing on each cycle a random outgoing transition from the current state.
For complete FSMs, each transition was prohibited to be executed with the probability of 50\% while generating scenarios.
The rationale behind this was to test the methods on instances for which the use of scenarios only is not sufficient to generate a proper FSM.
For each FSM we created 10 scenarios with a total length of $50|S|$.

While generating instances for the problem of incomplete FSM identification, we accepted only \emph{hard} instances: the ones for which the method from \cite{ulyantsev2012extended} is insufficient, i.e. it produces an FSM not compliant with the formulae.
If the instance was not hard, it was replaced with a new one until the new instance appeared to be hard.
A similar approach was applied in \cite{chivilikhin2014combining}.
Thus, we prepared satisfiable problem instances, for which a solution is guaranteed to exist.
The entire instance generation procedure is outlined in Algorithm~\ref{instance_preparation_algo}.

\begin{algorithm}
 \SetKwData{FSM}{FSM}\SetKwData{CheckFSM}{CheckFSM}\SetKwData{SC}{SC}\SetKwData{LTL}{LTL}
 \SetKwData{complete}{complete}\SetKwData{numb}{number}
 \SetKwFunction{Generate}{RandomFSM}\SetKwFunction{GenSC}{RandomScenarios}\SetKwFunction{GenLTL}{RandomLTL}
 \For{$|S| \in 3..12$, \complete $\in$ \{true, false\}, $i \in 1..50$}{
   \Repeat{\upshape the method from \cite{ulyantsev2012extended} run on $\FSM_i$ and $\SC_i$ \algindent produces an FSM incompliant with $\LTL_{i, 1..4}$}{
     $\FSM_i$ $\leftarrow$ \Generate{$|S|$, $|E| = 4$, $|Z| = 4$, \algindent \complete}\\
     $\SC_i$ $\leftarrow$ \GenSC{$\FSM_i$, $\numb = 10$, \algindent $l = 50|S|$, \complete}\\
     \For{$j \in 1..4$}{
       \Repeat{$\LTL_{i, j}$ \upshape is true for less than 6 FSMs \algindent out of $\CheckFSM_{1..10}$}{
         $\LTL_{i, j}$ $\leftarrow$ \GenLTL{$\FSM_i$}\\
         \For{$k \in 1..10$}{
           $\CheckFSM_k$ $\leftarrow$ \Generate{$|S|$, \algindent $|E| = 4$, $|Z| = 4$, \complete}\\
         }
       }
     }
   }
 }
 \Return{$\left(\FSM_{1..50}, \SC_{1..50}, \LTL_{1..50, 1..4}\right)$}\\
 ~\\
\caption{Instance generation procedure.}
\label{instance_preparation_algo}
\end{algorithm}

\subsubsection{Experiment setup}
Similarly to the experiments with the case study systems, we again aimed to find a minimum FSM for each instance and thus applied the same procedure of iterating over $|S|$ as before.
Note that the number of states $|S|_{\min}$ of a minimum FSM can not exceed $|S|_{\max}$ of the FSM from which the scenarios were generated.

While evaluating the methods, we wished to determine how many instances could be solved by the methods within a reasonable time limit.
The time span given for each method to solve each $|S|$-iteration of each instance was chosen to be 5 minutes.
If either of the iterations failed, then the whole run for the instance was regarded as failed.
Except violating the time limit, the Exponential SAT-based method could fail due to the lack of memory for its operation.

For each problem subtype (complete and incomplete FSM identification), $3 \varleq |S|_{\max} \varleq 12$ and each FSM construction algorithm, 50 executions were performed for target FSMs with these properties.
The number of solved instances was recorded for each set of executions.

\subsubsection{Results}
\label{sec:results}
Table~\ref{tab:succ} shows the results of FSM identification method executions in terms of numbers of solved instances.
It is clearly visible from the table that the QSAT-based approach again performed almost inadequate.
Following that, the Exponential SAT-based solution solved the majority of instances being unsuccessful only on incomplete instances with large $|S|_{\max}$.
Finally, the leaders among the strategies were the Backtracking and the Iterative approaches with the latter being better on large $|S|_{\max}$.
The obtained results are consistent with the ones obtained in Section~\ref{sec:exp_eval_case}.

\begin{table*}
\centering
\caption{Numbers of problem instances (out of 50) solved by FSM identification method executions arranged by the type of problem (complete and incomplete identification) and the number of states $|S|_\mathrm{max}$.}
\begin{tabular}{c@{\hspace{0cm}}ccccc@{\hspace{0.15cm}}ccccc}
\hline\myskip
\multirow{2}{*}{$|S|_\mathrm{max}$} && \multicolumn{4}{c}{Complete FSM identification} && \multicolumn{4}{c}{Incomplete FSM identification}\\\cline{3-6}\cline{8-11}\myskip
   && ITER & QSAT & EXP & BTR && ITER & QSAT & EXP & BTR \\\hline\myskip
3  && 50   & 47   & 50  & 50  && 50   & 42   & 50  & 50  \\
4  && 50   & 34   & 50  & 50  && 50   & 19   & 50  & 50  \\
5  && 50   & 33   & 50  & 50  && 50   & 12   & 50  & 50  \\
6  && 50   & 23   & 49  & 50  && 50   & 6    & 45  & 50  \\
7  && 50   & 20   & 47  & 50  && 50   & 5    & 37  & 50  \\
8  && 50   & 20   & 46  & 47  && 50   & 4    & 28  & 48  \\
9  && 50   & 11   & 48  & 44  && 50   & 1    & 17  & 42  \\
10 && 50   & 10   & 46  & 41  && 46   & 0    & 19  & 28  \\
11 && 50   & 9    & 41  & 38  && 44   & 0    & 21  & 20  \\
12 && 50   & 15   & 44  & 30  && 43   & 0    & 18  & 7   \\
\hline
\end{tabular}
\label{tab:succ}
\end{table*}

We note that the complete instance set was generally easier for all methods to solve.
This was due to the fact that the minimum numbers $|S|_{\min}$ for this instance set were often less than the ones of the incomplete set.
For example, for $|S|_{\max} = 10$, FSMs had average $|S|_{\min}$ of 5.8 for complete instances and 8.5 for incomplete ones.

\begin{figure}
\centering
\includegraphics[width=3.32in]{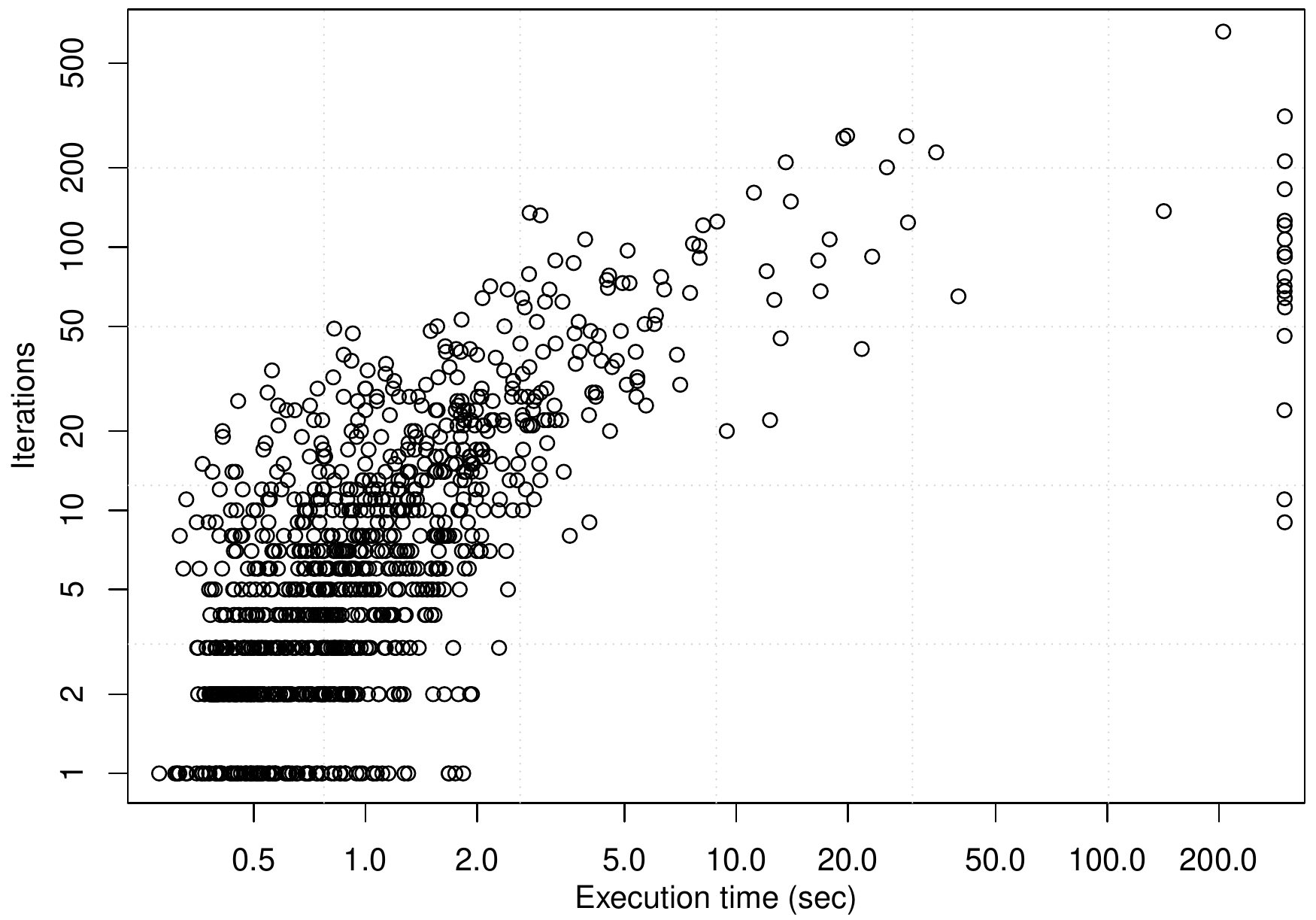}
\caption{Execution time vs. iteration number for the Iterative SAT-based method.}
\label{fig:plot_run_time}
\end{figure}

In addition to performance measurements, we examined in more detail the execution of the Iterative SAT-based approach, which is the most appealing one according on the results.
In Fig.~\ref{fig:plot_run_time}, its execution times are compared with iteration numbers (the data for all instances is combined).
From this plot it is clearly visible that a considerable fraction of instances (more precisely, 96.2\%) was solved within only 10 seconds.
The vertical row of point on the right corresponds to unsolved instances (1.7\%).
This plot additionally provides a glance at instance complexity: for example, 59.7\% of instances required at least five iterations to be solved.
Since the instances were filtered based on complexity during their generation, the number of instances solved in one iteration is small (8.6\%), which implies that a method not supporting LTL properties, like the one from \cite{ulyantsev2012extended}, would not perform adequately on our dataset.

\subsection{Comparison with inexact methods}
\label{sec:comparison}
This subsection reports on the comparison of the proposed FSM identification techniques with three inexact methods: the \muaco~algorithm  \cite{chivilikhin2014combining}, the state merging approach \cite{walkinshaw2008inferring}, and the \emph{Unbeast} tool \cite{ehlers2011unbeast} based on bounded LTL synthesis.
Since the performance of the QSAT-based method has been shown to be inappropriate, this method was excluded from experiments in this subsection.

\subsubsection{Comparison with the metaheuristic FSM identification method}
\label{sec:comparison_muaco}
We compared the Iterative SAT-based, the Backtracking and the Exponential SAT solutions with the \muaco~method introduced in \cite{chivilikhin2014combining} to infer FSMs with no completeness requirement.
In \cite{chivilikhin2014combining}, the evaluation of the method for three total scenario lengths ($l \in \{50|S|$, $100|S|$, $200|S|\}$) and six different state numbers ($5 \varleq |S| \varleq 10$) is reported.
For each combination of these parameters, 50 instances were prepared, and only hard ones were retained.
This test data was obtained from \cite{chivilikhin2014combining} with precise values of median method execution times (these values were shown in \cite{chivilikhin2014combining} on charts) and with annotations of which instances are hard.

Since the FSM model of \cite{chivilikhin2014combining} uses input variables, we transformed them into events as described in Section~\ref{sec:problem_statement}.
Problem instances considered in \cite{chivilikhin2014combining} have two variables and two events.
Hence, we end up with $|E| = 8$ new events.
Such a transformation also duplicates some edges in the scenario tree, since there might be several satisfying assignments for a Boolean guard condition of a scenario element.
Thus, we had to modify the Backtracking approach to handle multiple edges: each edge group originating from a common scenario element is always added or removed simultaneously.

Table~\ref{tab:comp} presents the comparison of median execution times of the proposed methods and the \muaco~algorithm on this data.
This time, since \muaco~is not an exact method, we did not try to minimize $|S|$ and executed the methods for the maximum possible numbers of states.
Note that \cite{chivilikhin2014combining} used the \emph{AMD Phenom II x}4 955 3.2 GHz CPU for experiments.
As for the proposed methods, which were still executed on an \textit{Intel Core i}7-4510{U} 2.0 GHz CPU, they were now given 15 minutes for $l = 50|S|$, 30 minutes for $l = 100|S|$ and 60 minutes for $l = 200|S|$, since solver execution time was affected by the length of scenarios and we wished the methods to solve the majority of instances.

\begin{table*}
\centering
\caption{Median execution times (in seconds) of the proposed methods and \muaco~(designated as CMA) on the instance sets from \cite{chivilikhin2014combining}. Asterisks (*) show lower bounds on medians in the cases when less than 50\% of runs were finished within their time limits.}
\begin{tabular}{c@{\hspace{0.1cm}}ccccccccccccccc}
\hline\myskip
\multirow{2}{*}{$|S|$} && \multicolumn{4}{c}{$l = 50|S|$}  && \multicolumn{4}{c}{$l = 100|S|$} && \multicolumn{4}{c}{$l = 200|S|$}\\\cline{3-6}\cline{8-11}\cline{13-16}\myskip
   && ITER & EXP  & BTR  & CMA   && ITER & EXP  & BTR   & CMA    && ITER & EXP   & BTR    & CMA    \\\hline\myskip
5  && 0.7  & 3.5  & 0.8  & 25.5  && 1.1  & 8.1  & 1.1   & 27.5   && 2.1  & 20.1  & 0.9    & 76.5   \\
6  && 0.9  & 6.1  & 3.8  & 32.0  && 1.5  & 15.6 & 10.2  & 19.0   && 2.8  & 40.2  & 4.1    & 41.5   \\
7  && 1.0  & 9.4  & 37.4 & 68.0  && 1.8  & 28.5 & 26.6  & 103.0  && 4.0  & 91.1  & 16.1   & 169.5  \\
8  && 1.2  & 10.8 & 900* & 410.0 && 2.5  & 27.9 & 465.1 & 191.0  && 5.3  & 61.2  & 3205.6 & 462.0  \\
9  && 1.4  & 39.1 & 900* & 213.0 && 3.0  & 41.9 & 1800* & 406.5  && 8.2  & 169.6 & 3600*  & 299.5  \\
10 && 2.4  & 15.9 & 900* & 958.0 && 3.9  & 40.9 & 1800* & 2443.0 && 9.9  & 408.8 & 3600*  & 4025.5 \\
\hline
\end{tabular}
\label{tab:comp}
\end{table*}

The Iterative approach solved all the instances, unlike the other methods, and, according to the table, its performance was also the best.
Next, the Exponential SAT-based approach failed to solve around 16\% of instances (92 of 580), mostly due to reaching the memory limit of 8GB (this can be explained by the fact that the length of the formula to be solved grows exponentially with $k$).
Still, this did not influence the execution time medians, and this approach outperformed \muaco~as well.
The superiority of the Iterative and the Exponential SAT-based approaches in comparison with \muaco~is most evident for $|S| = 10$.
Finally, the Backtracking approach was less successful: starting from $|S| = 8$, it generally lacked time to solve the instances.
Nonetheless, this approach has the benefit of being quite indifferent to scenario length and consumes little memory.

\subsubsection{Comparison with counterexample-based state merging}
\label{sec:comparison_sm}
A more narrow problem than the one considered in this paper is the problem of finite-state model identification stated in \cite{walkinshaw2008inferring}.
Their task was to construct an FSM whose transitions are marked only with events using a number of software execution traces (event sequences) and temporal properties, represented as LTL safety formulae.
An LTL formula is a safety formula, if every counterexample to it has a finite prefix such that every its continuation is a counterexample~-- such finite counterexamples were previously mentioned in Section~\ref{sec:negsctree}.
Informally speaking, such formulae ensure that some undesired conditions never become true.

The passive approach from \cite{walkinshaw2008inferring} (this paper also suggested the active approach, in which the user is inquired during FSM synthesis) was implemented independently, and instead of the Spin\footnote{\url{http://spinroot.com/}} verifier we used the one mentioned previously.
Following \cite{walkinshaw2008inferring}, we chose Blue Fringe \cite{lang1998results} as the state merging method.

As the data for comparison, we adopted the examples from \cite{walkinshaw2008inferring} with few changes.
In particular, we ensured that the data allowed the compared methods to produce the desired FSMs, if they were able to produce any FSMs at all.
The data used in this comparison has already been applied in Section~\ref{sec:exp_eval_case} for the case study evaluation.
It includes the instances in which the finite-state models of a simple text editor, the JHotDraw drawing framework and the Jakarta Commons CVS client are to be learnt.

The results of the method comparison are presented in Table~\ref{tab:walkinshaw}, which partially duplicates the data from Table~\ref{tab:case_results} in Section~\ref{sec:exp_eval_case}.
The performance of the proposed methods on the considered instances has been discussed previously.
As for state merging, it appeared to be much faster than the exact methods.
Nonetheless, although it identified target minimum FSMs, in general it does not guarantee the minimality of its solutions: it has no means of proving that its answers are optimal with respect to the number of states.
Moreover, as far as we know, there are no ways of applying it in the cases of FSMs with actions or non-safety (i.e. general) LTL properties.

\begin{table}
\centering
\caption{Execution times (in seconds) of the proposed methods and the passive state merging (SM) approach \cite{walkinshaw2008inferring} on three instances adopted from the same work. ML and TL stand for a failure due to the memory limit (8GB) and the time limit (48 hours) respectively.}
\begin{tabular}{cccccc}
\hline\myskip
Instance    & $|S|_{\min{}}$ & ITER & EXP & BTR  & SM   \\\hline\myskip
Text editor & 4              & 1.3  & 8.9 & 1.0  & 0.4  \\
JHotDraw    & 7              & 3.6  & ML  & 27.9 & 0.8  \\
CVS client  & 18             & TL   & ML  & TL   & 36.2 \\\hline
\end{tabular}
\label{tab:walkinshaw}
\end{table}

To demonstrate that state merging may not find a minimum FSM, we randomly generated 100 FSMs without output actions, each with 5 states, 10 events and 25\% of possible transitions.
To ensure that LTL formulae were safety ones, instead of using \emph{randltl} we generated them according to several simple templates.
One of such templates was $\temp{G}(e \to \temp{X}(e_1 \lor ... \lor e_k))$, where $e$ is an event and $e_1, ..., e_k$ are the events which can occur on a transition which follows a transition marked with $e$.
For 12 of 100 FSMs state merging failed to find the minimum answer.

Similarly to the approach from \cite{walkinshaw2008inferring}, our Iterative, Backtracking and Exponential SAT-based approaches can be modified to be active: while at least two FSMs can be identified from the input data, it may ask the user to confirm or reject an execution trace of one of such FSMs to obtain more data.
These approaches can also be applied to find all possible solutions of the problem.
To do this for solver-based methods, after a solution is found, a constraint prohibiting the found FSM is appended to the formula, and the solver is restarted.
Instead of this, the Backtracking approach can simply continue its search.

\subsubsection{Comparison with LTL synthesis}
Recent advances in LTL synthesis resulted in software tools called \emph{Lily} \cite{jobstmann2006optimizations}, \emph{Acacia} \cite{filiot2009antichain} and \emph{Unbeast} \cite{ehlers2011unbeast}.
Among them, we used \emph{Unbeast} for comparison since it is more recent and is claimed to be superior over others \cite{ehlers2012symbolic}.
Another tool \emph{G4LTL-ST} \cite{cheng2014g4ltl} is also known, but it is focused solely on program synthesis for PLCs and more rich forms of LTL specification.

Since in the problem of LTL synthesis the specification is given only as LTL properties, we had to encode scenarios in LTL.
While doing so we unfortunately lost the ability to distinguish the order of actions produced on the same cycle, which could potentially allow smaller FSMs to satisfy scenarios.

Next, the problem of LTL synthesis requires the construction of complete reactive systems.
We tried to express incompleteness using a dummy action for transitions which are to be removed after the synthesis and exempting paths which included such dummy transitions (i.e. actually impossible paths) from the need to match LTL properties.
Unfortunately, this led to the problem of states with no proper outgoing transitions; paths leading to such states were also exempted from compliance with temporal specifications.
Thus, we report on the results of running \emph{Unbeast} only on instances for complete FSM identification.

\emph{Unbeast} produces the target system as a game which resembles a Mealy FSM.
Using the provided game simulator, it is possible to reconstruct this FSM.
After the reconstruction, we also greedily minimized the FSM maintaining the compliance with the specification.

We executed \emph{Unbeast} on complete instances from our instance set described in Section~\ref{sec:experiment_setup} for $3 \varleq$ $|S|_{\max{}}$ $\varleq 6$ (for larger $|S|_{\max{}}$ the execution time of \emph{Unbeast} was often impractically large).
Executions for various $|S|_\mathrm{max}$resulted in median run times of 8, 38, 130 and 510 seconds, respectively.
As described in Section~\ref{sec:results}, the majority of the methods proposed in this paper performed better.
Furthermore, the number of states in FSMs produced by \emph{Unbeast} was immense, reaching the median of 103 already for $|S|_\mathrm{max} = 3$.

The reason for the weak performance of $\emph{Unbeast}$ seems to be the large size of LTL specifications, which mainly results from scenarios.
Shortened scenarios were noticed to cause better performance.
We conclude that while $\emph{Unbeast}$ might perform well on natural LTL specifications, the inclusion of scenarios, which are typical for the problem of software model reverse engineering, causes it to perform much worse.

\subsection{Threats to validity}
All three groups of experiments, which are described above, are potentially prone to validity threats.
In Section \ref{sec:exp_eval_case}, a number of instances from the literature are used to evaluate the proposed methods.
This set of instances is small, and hence might be not representative with respect to other software models.
This threat has been partially mitigated by selecting case instances from different studies.

Next, the random FSM generation procedure (Algorithm~\ref{instance_preparation_algo}) might have produced nonsensical or oversimplified reference models.
Unfortunately, the extent to which randomly generated models are similar to software models encountered in practice is problematic to determine, so threat mitigation was limited to instance complexity evaluation.
First, the data generation algorithm (Algorithm~\ref{instance_preparation_algo}) was equipped with complexity filtering.
The numbers of iteration required to solve the remaining instances were provided in Fig.~\ref{fig:plot_run_time} and appear to be satisfactory.
Next, a subset of generated models was examined visually, which confirmed that state transition graphs of the generated FSMs were sufficiently complex.
Following that, minimum solution sizes were determined during the evaluation.
According to Section~\ref{sec:results}, the complete instance set was generally easier in terms of this measure; nevertheless, instances with $|S|_{\max} = 12$ were able to distinguish the performance of different methods.

A possible error in the implementations of methods might have made the corresponding execution metrics meaningless.
To prevent this, for each solved instances its correspondence with the required specification was ensured.
Finally, in Section~\ref{sec:comparison_sm}, we use our own implementation of the state merging method from \cite{walkinshaw2008inferring}, which might perform differently than in \cite{walkinshaw2008inferring}.
Nevertheless, the performance of this method is clearly superior to the one of the proposed ones, and the shortcomings of this approach are mainly qualitative and do not depend on the implementation.

\subsection{Discussion}
\label{sec:discussion}
It remains to discuss the results from all subsections of the evaluation and finally answer the research questions formulated in the beginning of Section~\ref{sec:exp_eval}.
As revealed in Section~\ref{sec:case_results}, the Iterative SAT-based and the Backtracking methods were able to cope with the majority of considered case instances, which answers RQ1~(a).
However, it was also found that for large numbers of states the performance of these methods (especially the one of the Iterative SAT-based method) could be much better if the optimal number of states was known in advance.
This raises the question whether the precision of solving the problem can be partially traded for performance.
This question is discussed in more detail in Section~\ref{sec:conclusion}.

With respect to scalability, which is questioned in RQ2, the methods are ranked in the following order: the Iterative SAT-based, the Exponential SAT-based, the Backtracking, and the QSAT-based methods.
These are the results of both Section~\ref{sec:results} are Section~\ref{sec:comparison_muaco}.
The first of these methods was able to solve almost all of the hardest instances considered.

As for RQ3, the comparison of the proposed methods with the known ones can be viewed in terms of capabilities and performance.
In terms of capabilities, the considered alternative methods lacked some of the features supported by the proposed ones.
However, we did not compare our methods with the ones which are able to synthesize richer finite-state models, like in \cite{ohmann2014behavioral,walkinshaw2016inferring}, which is obviously an advantage of such methods.
The results in terms of performance are quite diverse.
Even the most scalable Iterative SAT-based method did not excel state merging, but one should remember that the latter is quite limited in the set of problem instances it can handle (this is also the reason why state merging was not included in the majority of experiments considered in this paper).
On the other hand, two of the proposed methods were able to surpass \muaco~(see Section~\ref{sec:comparison_muaco}).

Finally, we must answer RQ1~(b) by determining whether the proposed methods are potentially applicable in industrial software engineering.
All empirical evaluations performed in this study assume that the number of FSM states is quite small: it did not exceed 18.
Clearly, this number of states is insufficient to represent real-world systems in full detail.
Nevertheless, identified models should not be considered as full-scale substitutes of the target systems.
Their purpose is to provide a picture of the system's logic for a software engineer, which becomes more clear and concise once the minimality requirement is fulfilled.
This picture can serve as a basis either for understanding or reverse-engineering the system.
It is also worth noting that FSMs which include 20 or more states are difficult to comprehend.
If this number of states is insufficient for the given specification, one might try simplifying it by focusing only on particular aspects of the system.
These thoughts, as well as the performed case study evaluation, make us hope that the Iterative SAT-based approach (the proposed approach with the best performance) is potentially applicable in practice, although more research is needed to improve it.

Aside from research questions, we need to explain why the methods obtained the results observed in the evaluation.
Despite complex theory behind, the QSAT-based approach is clearly an outsider.
This might be due to the low performance of the state-of-the-art QSAT solvers, which will hopefully be improved in the future.
Another possible reason is the consideration of only one translation \cite{biere2003bounded} of the BMC problem into SAT.
Some other translations can be found in \cite{amla2005analysis}.

The SAT-based implementation of the QSAT-based approach, the Exponential SAT-based approach, performed better, but at least two drawbacks still hindered its performance.
First, the length of the BMC part of the Boolean formula is exponential of the LTL property length in the worst case.
Second, the length of the formula is exponential of $k$, therefore in our experiments the tractable value of $k$ did not exceed four.
This means that only counterexamples with length up to five were taken into account, which is clearly insufficient for large FSMs.
When a larger value of $k$ was required, this method consumed too much memory.

Next, the good results of the Backtracking approach were quite surprising given its simplicity.
Because of it, it easily solves small instances, but its performance drops fast when the number of states becomes sufficiently high (around 11 in Table~\ref{tab:succ} and around 8 in Table~\ref{tab:comp}).
A possible reason why this approach performs worse in Table~\ref{tab:comp} compared to Table~\ref{tab:succ} is the more complex structure of the scenario tree (see Section~\ref{sec:comparison_muaco}).
The Backtracking approach is also not prone to memory problems, unlike the Exponential SAT-based approach.

Finally, possible reasons for the success of the Iterative SAT-based approach include matching the idea of the method with the efficient incremental solver \emph{cryptominisat}, and a relatively small overhead of encoding counterexamples in the Boolean formula compared to the Exponential SAT-based approach.
We must note that the performance of this approach can be much worse on large unsatisfiable problem instances (see Section~\ref{sec:case_results}).

\section{Conclusions and future work}
\label{sec:conclusion}

We have presented a number of approaches for the exact FSM identification problem from scenarios and temporal properties.
This problem arises when one wants to infer or reverse engineer the model of a software system, such as a desktop application or an industrial reactive system (e.g. elevator controller).
The approaches were evaluated and compared with existing inexact (i.e. not producing minimum FSMs) solutions both on case study instances from previous research and on randomly generated instances.
The ranking of the methods in terms of their performance, from the best to the worst, was determined to be as follows: the Iterative SAT-based, the Exponential SAT-based, the Backtracking, and the QSAT-based approaches.

Unlike the earlier metaheuristic approach \cite{chivilikhin2014combining}, the proposed ones support the FSM completeness requirement and are able to prove the unsatisfiability of instances, which can be applied to construct minimum FSMs compliant with the given specification.
However, on a more narrow problem of identifying FSMs without actions from scenarios and LTL safety properties, the proposed approaches are outperformed by the one from \cite{walkinshaw2008inferring}.
As for the comparison with the symbolic bounded LTL synthesis tool \cite{ehlers2011unbeast}, it reveals that scenarios, which usually do not cause problems for the presented approaches, make this tool produce large solutions and work slowly.

The performance of the Iterative SAT-based method is encouraging, but supporting larger FSMs would bring it closer to industrial application.
An alternative way of applying FSM synthesis in software engineering is sacrificing features, such as precision and full support of LTL, in order to improve performance.
If only positive examples are used for learning, then prohibiting invalid model behaviors is only possible by providing an exhaustive set of scenarios.
Also, as shown in \cite{walkinshaw2008inferring,lo2009automatic,beschastnikh2011leveraging}, temporal properties are important in guiding FSM synthesis, thus the entire refusal of LTL specifications would be discouraging.
Among known methods, a good option is the method from \cite{walkinshaw2008inferring}.
However, it supports only the safety subset of LTL, which, for example, does not include the common unbounded response property $\temp{G}(x \to \temp{F}y)$.
As for the methods proposed in this paper, Section~\ref{sec:case_results} has shown that proving the optimality of the solution can be much more time consuming than finding the solution.
Thus, finding the minimum solution approximately may be a proper trade-off between precision and performance.

Another possible future work direction might involve considering other translations from BMC into SAT \cite{amla2005analysis}, which may improve the QSAT-based solution.
Then, the methods can be brought closer to practice by supporting wider classes of FSMs, like in \cite{ohmann2014behavioral} and \cite{walkinshaw2016inferring}.
One more interesting idea to try is automatic mining of temporal properties from execution traces \cite{lo2009automatic,beschastnikh2011leveraging}, which can solve the difficulty of obtaining temporal specifications.

\bibliographystyle{spmpsci}
\bibliography{ulyantsev-sttt}\vspace{0cm}
\end{document}